\documentclass[modern]{aastex61}

\usepackage{amsmath}
\usepackage[flushleft]{threeparttable}

\shorttitle{FRB 121102 Detection and Periodicity}
\shortauthors{Zhang et al.}

\begin{document}

\title{Fast Radio Burst 121102 Pulse Detection and Periodicity: \\
A Machine Learning Approach}

\author{Yunfan Gerry Zhang}
\email{yunfanz@berkeley.edu}
\affil{Department of Astronomy, University of California Berkeley}
\affil{Berkeley SETI Research Center, University of California, Berkeley}
\author{Vishal Gajjar}
\affil{Space Sciences Laboratory, Berkeley, California}
\affil{Berkeley SETI Research Center, University of California, Berkeley}
\author{Griffin Foster}
\affil{Sub-Department of Astrophysics, Oxford University}
\affil{Department of Astronomy, University of California Berkeley}
\affil{Berkeley SETI Research Center, University of California, Berkeley}
\author{Andrew Siemion}
\affil{Department of Astronomy, University of California Berkeley}
\affil{Berkeley SETI Research Center, University of California, Berkeley}
\affil{SETI Institute, Mountain View, California}
\affil{Radboud University, Nijmegen, Netherlands}
\affil{Institute of Space Sciences and Astronomy, University of Malta}
\author{James Cordes}
\affil{Department of Astronomy, Cornell University}
\author{Casey Law}
\affil{Department of Astronomy, University of California Berkeley}
\author{Yu Wang}
\affil{Department of Statistics, University of California Berkeley}


\begin{abstract}

We report the detection of 72 new pulses from the repeating fast radio burst FRB 121102 in Breakthrough Listen C-band (4-8~GHz) observations at the Green Bank Telescope. The new pulses were found with a convolutional neural network in data taken on August 26, 2017, where 21 bursts have been previously detected. Our technique combines neural network detection with dedispersion verification. For the current application we demonstrate its advantage over a traditional brute-force dedispersion algorithm in terms of higher sensitivity, lower false positive rates, and faster computational speed. Together with the 21 previously reported pulses, this observation marks the highest number of FRB 121102 pulses from a single observation, totaling 93 pulses in five hours, including 45 pulses within the first 30 minutes. The number of data points reveal trends in pulse fluence, pulse detection rate, and pulse frequency structure. We introduce a new periodicity search technique, based on the Rayleigh test, to analyze the time of arrivals, with which we exclude with 99\% confidence periodicity in time of arrivals with periods larger than $5.1$ times the model-dependent time-stamp uncertainty. In particular, we rule out constant periods $\gtrsim10$\,ms in the barycentric arrival times, though intrinsic periodicity in the time of emission remains plausible. 

\end{abstract}




\section{Introduction} \label{sec:intro}

Fast Radio Bursts (FRBs) are millisecond-duration radio transients that exhibit
dispersion relations consistent with propagation through cold plasma
\citep{Lor07,Tho13,Pet16}. Out of the known FRBs, only FRB 121102 has been observed
to repeat \citep{Spi14, Spi16, Sch16, Sch17}. The repeating pulses allowed
precise localization of the source within a dwarf galaxy of redshift 0.193
\citep{Cha17,Mar17,Ten17} confirming the extra-galactic nature of the phenomenon
suspected from their high dispersion measures. Recently, Breakthrough Listen
reported observations of 21 pulses of FRB 121102 recorded with the C-band
receiver at the Green Bank Telescope (GBT; \citealt{vishal}). The reported bursts
marks the highest frequencies of pulses ever detected from the
repeating FRB. Together with the observation at the William E. Gordon Telescope
at the Arecibo Observatory, the new pulses showed   100\% linearly polarized
emission with high and variable rotation measure of $+1.33\times 10^5$ radians
per square meter to $+1.46\times 10^5$ radians per square meter in the source reference frame, indicating that
the source is situated in a highly magneto-ionic environment
\citep{Michilli:2018}. 

In this paper, we present a re-analysis of the C-band observation by
Breakthrough Listen on August 26, 2017 with convolutional neural networks.
Recent rapid development of deep learning, and in particular, convolutional
neural networks (CNN; \citealt{AlexNet,VGG,resnet,Inception}) has enabled revolutionary
improvements to signal classification, pattern recognition in all fields of data
science such as, but not limited to computer image processing, medicine, and
autonomous driving. In this work, we present the first successful application of deep learning to direct detection of fast radio transient signals in raw spectrogram data. Deep learning methods have been applied to pulsar search in \cite{zhu2014,guo2017}, while \cite{Wags2016,Foster2018} apply traditional machine learning to single pulse transient candidate classification. Recently \cite{connor18} applies deep learning models to FRB searches. These work all focus on reducing false positive rate from candidates of traditional search, though \cite{connor18} suggests the possibility of direct deep learning detection. As we shall see, neural networks can in some scenarios offer higher sensitivity, but lack interpretability in their predictions. Dedispersion searches are interpretable but may suffer from poor sensitivity false-alarm trade-off.
In this work, we leverage the advantages of both techniques by using the latter to
verify the candidate outputs of the former. The resulting technique revealed
more than 70 new pulses of FRB\,121102 in a 5 hour C-band observation conducted
by Breakthrough Listen. Our neural network is capable of processing Breakthrough Listen spectral-temporal data 70 times faster than real time on a single GPU, though processing speed in other contexts depends on the frequency and time resolution. We do not claim our technique is ready to replace current state of the art dedispersion pipelines, but our method shows advantage in some scenarios and encourages further exploration.

The unprecedented rate of pulses from the repeating FRB allows us to explore trends in the pulse fluence, arrival times, and frequency structure. In particular, an abundant list of theoretical models take invested interest in the search for periodicity in the pulse arrival times. Previous works have reported non-detection of periods in the pulses \citep{Spi14,Spi16,Sch16}, though none have been able to quantify the statistical significance of the null detection. In this work, we introduce a new method of period detection that is highly sensitive given the limited number of pulses. With this method we are able to exclude all periods in the arrival times $\gtrsim10$~ms with 99\% confidence. 

The contributions of this paper thus include:

1. The first successful application of deep learning to direct detection of fast radio transient signals, with  72 new pulses of FRB\,121102. 

2. Trends of pulses structure from FRB\,121102 with the most number of pulses detected from a single observation to date. 

3. A new periodicity search technique applied to repeating radio transient detections. 

4. A first statistical limit of aperiodicity on the detected pulses of FRB\,121102. 

The rest of this paper is organized as follows. In Section \ref{sec:observation} we describe the Breakthrough Listen C-band observation and digital back-end at the Green Bank Telescope. In Section \ref{sec:method} we describe the detection method with a convolutional neural network and verification with dedispersion. In Section \ref{sec:properties} we describe a list of properties of the pulses and the procedures we use to determine them. In Section \ref{sec:discussion} we discuss the trends in these properties, as well as limits on periodicity. In Section \ref{sec:conclusion} we conclude.

\section{Observation} \label{sec:observation}
The observations that produced the detections described here were conducted as a component of the Breakthrough Listen Initiative (BL), a comprehensive search for extraterrestrial intelligence currently employing a number of radio and optical observatories \citep{wds+17}.  The bulk of the targets of the BL observational program consists of nearby stars, nearby galaxies and the Milky Way galactic plane, but a small component of the program includes ``exotica'' \citep{ism+17}.   This latter category includes a variety of targets, including anomalous astronomical sources with a potential relationship to Extra-terrestrial Intelligence (ETI).  Following a number of suggestions in the literature that FRBs exhibited some characteristics that we might expect from an ETI transmitter (e.g. \citealt{Loeb17}), known FRB sources, including FRB 121102, were added to the BL observing queue.  

On August 26, 2017, we initiated a 6 hour observing session of FRB 121102 using the 4$-$8~GHz (C-band) receiver on the GBT. This session ultimately yielded ten 30-minute scans of FRB 121102. Observations were conducted with the BL digital back-end \citep{macmahonaccepted}, which generates both time domain voltage data, and integrated spectral data. The spectral-temporal filterbank data used in this analysis have time and frequency resolution of 350~$\mu$sec and 366~kHz respectively. 

\section{Detection} \label{sec:method}
\subsection{Overview} \label{sec:overview}
De-dispersion based single pulse algorithms such as HEIMDALL \citep{Bar12}, FDMT \citep{FDMT}, Bonsai \citep{bonsai}, Amber \citep{Amber} and
CDMT \citep{cdmt} allow signal detection by re-aligning the frequencies according to
series of dispersion measure (DM), summing over frequency to get a time series, and thresholding the result. These algorithms face challenges of noise and radio frequency
interference (RFI) masquerading as false positives. To this end, the frequency-integrated signal to
noise ratio (S/N) of reported FRB pulses often extend down to an arbitrary detection
threshold \citep{KatzReview,Pet16}, implying a possibility of more pulses in archival datasets. Furthermore, as \cite{vishal} shows, in wide-band
observations the frequency structure of pulses is highly variable, which
additionally impairs the effectiveness of dedispersion-based algorithms when the de-dispersed data are summed over the entire bandwidth. The effect of
frequency structure can be mitigated by sub-banding the search \citep{bonsai,Amber,FDMT}, though an effective sub-band search or similar technique would need to be sufficiently flexible for the wide variety of observed pulse morphologies (e.g. \citealt{Spi2012}). In this work we deploy an alternative approach using deep learning.

Recently, deep convolutional neural networks drastically improved upon traditional machine learning methods of object detection with their ability to extract complex features in very large datasets. With their demonstrated ability to recognize objects within complex environments, convolutional neural networks are an excellent candidate for FRB searches. Current machine learning techniques can be broadly classified into supervised learning, unsupervised learning and reinforcement learning. Supervised object recognition is by far the most mature of the three and involves training a model to match decisions in a labeled ``training set". In the case of a deep convolutional neural network, the input is passed through multiple layers of convolutional and non-linear operations, the last of which outputs a probability of the decision. During training, the parameters of the convolutional kernels are updated to match the output probabilities to the ground truth labels for each sample in the training dataset. The trained model can then be used to recognize objects in a new dataset. 


\subsection{Data Preparation} \label{sec:data}
For pulse detection we use the standard high time resolution filterbank data produced by the Breakthrough Listen digital back-end \citep{macmahonaccepted}. The filterbank data are two dimensional spectrograms that span $4$~GHz to $8$~GHz with frequency resolution of 366kHz and a sampling time of $0.350$~ms. To create the
set of samples, we cut the data along the time axis into frames of 256 samples,
roughly capturing the dispersed duration of observed pulses. We find it
unnecessary to overlap the frames as the model demonstrated an ability to detect pulses close to the edge of the frames. To regularize over the bandpass we subtract the mean and divide by the standard deviation per channel, per frame \footnote{Averaging per frame, instead of batch of frames, reduces the influence of occasional very bright RFI.}. The resulting inputs are arrays of zero mean and unit variance. 

Supervised learning requires a labeled training set. In this case, we use two classes of samples: positive frames that contain at least one FRB pulse, and negative frames that contain only noise \footnote{As a clarification in terminology, radio astronomers sometimes use the word noise to mean both RFI and thermal pixel noise. In this paper we use the word exclusively in reference to pixel-noise in the spectrograms. } and RFI. Samples in the training set should ideally capture all the variations one would encounter in an inference scenario, including characteristics of FRB pulses, RFI, and background noise. Due to the rare occurrence of known FRB pulses, there are not enough real examples to capture all the variations in pulse characteristics (dispersion measure, amplitude, width, scintillation pattern, and location within the frame) for a search algorithm. However, the simplicity of the pulses allows relatively easy simulation. To this end, the scarcity of pulses becomes an advantage because there are abundantly available negative examples that capture the noise and RFI characteristics. Thus we choose to simulate pulses to superimpose on preprocessed negative examples. 

\begin{figure}
\begin{center}
\includegraphics[width=0.6\linewidth]{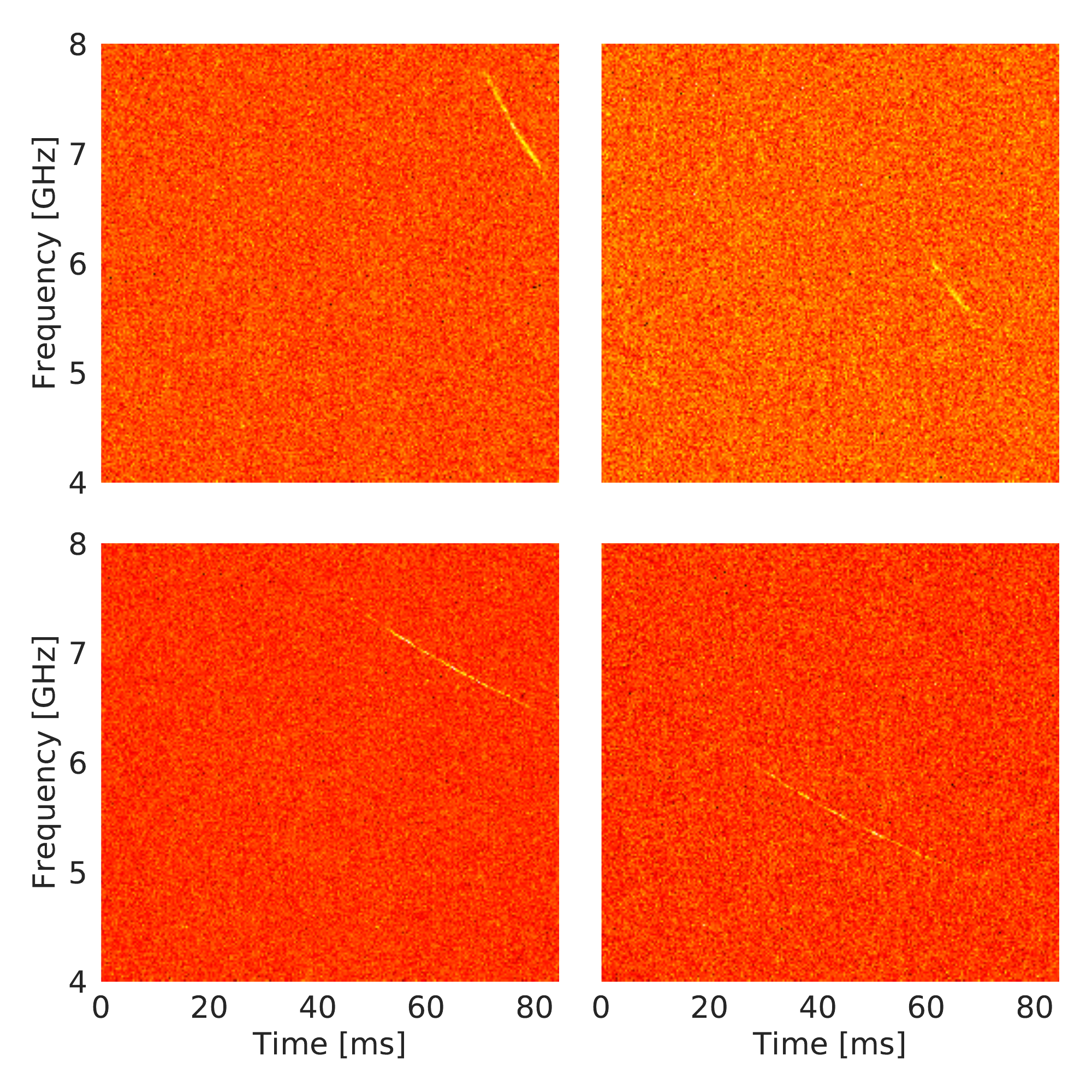}
\caption{Examples of simulated pulses on real observations. Relatively bright examples are shown for visual clarity while actual training set contains much weaker pulses. }
\label{fig:example_simu}
\end{center}
\end{figure}

For the training and test set, we use five hours of Breakthrough Listen observations with no detected pulses. For frames of 256 time samples each, this creates a training set of around 400000 images, half of which contain simulated pulses, the other half do not. Four and half hours of the observation is used to train the network, while the remaining half hour is set aside as an independent test set. The simulated pulses have dispersion measures sampled from uniform distribution from 200 to 2000~pc~cm$^{-3}$, as well as a wide range of bandwidth, pulse width and amplitude. To mimic the frequency modulation seen in reported detections \citep{vishal}, we add frequency modulation as a product of a Gaussian profile and a Markov mask\footnote{Specifically, we create an one dimensional binary frequency-channel mask using a Markov chain of constant transition probability. We then smooth the array by convolution with a one dimensional Gaussian kernel and apply onto the simulated pulse frequency profile to roughly emulate the observed frequency modulations in \cite{vishal}.}. In the time direction, we use Gaussians of the squares of time for the pulse profile to mimic the typical sharper-than-Gaussian drop-off. We vary the burst arrival time over a range of values within and outside the frame, ensuring that the pulse appears within the frame at their peak frequencies. Our simulation details are thus empirically motivated by the observed morphologies from this observation. Models of pulse morphologies based on astrophysical motivations are available. However we find our simple simulation sufficient for detection with a neural network. We show examples of simulated pulses in Fig. \ref{fig:example_simu}. The shown examples are relatively bright for visual clarity, while actual training set contains much weaker pulses.

\subsection{Model} \label{sec:model}
This section assumes basic familiarity with common terminology of CNNs. Reviewing the basics of CNN is beyond the scope of this paper. Readers unfamiliar with terminology such as ``activation function" and ``fully connected layer" may consult the abundantly available material online, such as the detailed Deep Learning Book \citep{deeplearningbook} or the concise glossary by \texttt{DeepLearning4j}\footnote{\href{https://deeplearning4j.org/glossary}{www.deeplearning4j.org/glossary}}. 

Spectrogram data from radio astronomy presents different challenges to common datasets in computer vision such as MNIST \citep{mnist} or Imagenet \citep{imagenet}. Compared to everyday objects such as cars or persons, the quadratic form of FRB pulses are relatively simple. Such low level features are thus typically captured in the early layers of a neural network. However, the network must have sufficient capacity to cope with the presence of high pixel-noise, especially since we are interested in signal strength down to and below the noise amplitude. 

Residual networks are a class of very deep convolutional neural networks proposed in \cite{resnet}. The central idea involves a ``skip-connection'' that shortcuts blocks of convolutional layers, which then computes only the ``residual" features. The skip connection was introduced to reduce overfitting, and thus allowing high capacity with much deeper networks. In our case, the skip connection also intuitively encourages propagation of low level features such as a quadratic burst to much deeper in the network. Many empirical and intuitive improvements have been proposed since the initial residual network. Here we employ an architecture similar to the model in \citep{WRN}, but with reduced width due to the expected smaller number of features. Table \ref{tab:architecture} shows the architecture of our network. The network is made up of stacks of convolutional blocks, which, with the exception of \texttt{conv0}, each consists of two sub-blocks linked by a dropout layer. The sub-blocks are convolution preceded by batch normalization and $relu$ activation unit, a design following the findings of \cite{preactiv}. Our network thus consists of 17 total convolutional layers and 6.2 million trainable parameters. With the exception of the last fully connected layer, we avoid pooling and use strided convolutions to reduce image size. To increase the signal to noise ratio and reduce the input complexity we collapse every 32 frequency channels in the first layer by a stride-32 convolution. Intuitively, we choose such a reduction factor such that unit resolution in frequency corresponds roughly (on the same order) as unit resolution in time for the dispersion relation of a FRB with DM in range of 100 to 1000~pc~cm$^{-3}$, as one can check by differentiating the dispersion relation at the C-band frequencies.  In practice, the resulting frequency resolution of 11.71MHz and time resolution of 0.35~ms is empirically sufficient for a wide range of pulse morphology, width and DM for the $4-8$~GHz band. 

\begin{table*}
\begin{center}
\begin{tabular}{c | c | c}
\hline
Group Name & Output Size & Stack Type \\ 
\hline
conv0 & $342\times256\times1$ & $[32\times 1]\times1$\\
conv1 & $171\times 128\times32$ & $[7\times 7]\times1$\\
conv2 & $42\times32\times32$ & $[3\times 3]\times2$\\
conv3 & $10\times8\times64$ & $[3\times 3]\times3$\\
conv4 & $5\times4\times128$ & $[3\times 3]\times2$\\
avg-pool & $1\times 128$ & {}\\
fc & 2 & {}\\
\hline
\end{tabular}
\caption{Architecture of our residual network, showing input, five convolutional stacks (conv), average pooling (avg-pool) and fully connected (fc) output layers. Stack types are shown in $[h\times w]\times N$, where $h$ and $w$ are the height and width of the weights, N is the number of convolutional blocks in the stack. The output sizes are shown as $[H\times W]\times M$, where M is the number of features. }
\label{tab:architecture}
\end{center}
\end{table*}

\subsection{Model Evaluation} \label{sec:finetune}
\begin{figure}
\begin{center}
\includegraphics[width=0.6\linewidth]{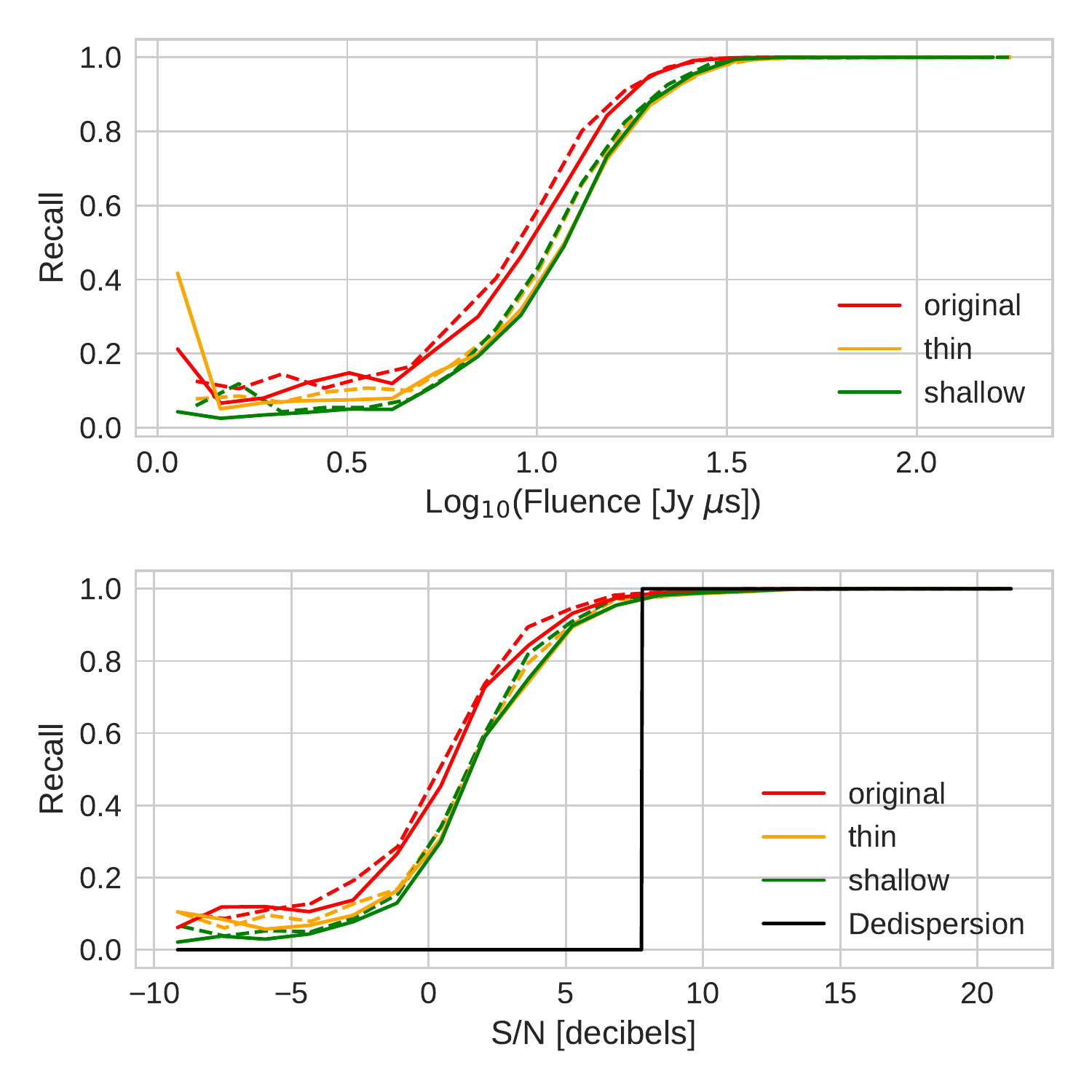}
\caption{Model recall scores as a function of fluence (top) and full-band frequency integrated S/N (bottom). The dashed lines show the training recall while the solid lines show test recall. Two smaller models are shown in comparison with our original (red). Compared to the original model, the thin model has half the amount of features and the shallow model has half as many layers. Both show reduced recall for weak pulses. The flare at the low fluence side is due to small-number statistics. In the lower panel, the theoretical upper limit recall for full-band dedispersion search with $6\sigma$ threshold is included for comparison. }
\label{fig:evaluation}
\end{center}
\end{figure}

The training of our \texttt{TensorFlow} model takes roughly 20 hours on a Nvidia Titan Xp GPU. Training error converges after 100 epochs to 93\%.  
We define two common terms for evaluation:
\begin{itemize}
\item Recall: The percentage of signals detected.
\item Precision: The percentage of detections that are real signals. 
\end{itemize}
Recall is thus a measure of the sensitivity of a model, while precision is a measure of robustness to false alarms. The overall recall for the test set is 88\%, and the precision is 98\%. We plot the average recall scores as a function of simulated pulse fluence and full-band frequency-integrated S/N in Fig. \ref{fig:evaluation}. The dash line shows the training recall, while solid line shows test-set recall. Typically, the training errors are lower than testing error, indicating a certain degree of over-fitting. In this case our model (red) shows only minimal degree of over-fitting. In the lower panel S/N is expressed in logarithmic scale:
\begin{equation}
\text{decibels} = 20\log_{10}\left(S/N\right). 
\end{equation}
In comparison, we show the theoretical upper limit for a full-band dedispersion search with a threshold of $6\sigma$. 

As we shall see in Section \ref{sec:discussion}, with three exceptions, our (sub-banded) dedispersion based verification procedure only yielded pulses with 20 Jy$\mu$s or larger. Our neural network model is able to detect these signals with over 95\% recall score, and is thus sufficient for detection of all verifiable pulses in this analysis\footnote{In fact, the fluences quoted in later sections are only lower bounds since we are unable to use the full band. The full band fluence are likely above 30 Jy$\mu$s for the weakest pulses, in which case our model has over 0.99 recall score.}. We also show two smaller models for comparison. With the ``thin'' model, we reduce the number of features of all layers by half, and with the ``shallow'' model we reduce the number of layers by half. Both of these models show noticeable drop in recall compared to the original. While it is certainly possible to obtain higher sensitivity with a larger model, here we choose an architecture that is sufficient for this analysis. 

 The accuracy of detection with a sufficiently capable network is determined by the quality of the training set. A good training set not only needs to be large enough to capture the distribution of inputs, but also be relatively balanced. Since our positive examples are simulated, we are able to have exactly balanced representation of positive signals, resulting in good recall score for all ranges of DM, width and frequency modulation. The RFI distribution, however, is not necessarily balanced. The rate of false detection in our network is around 2\%. Out of the 400000 images in our training set, if a type of RFI only exists in 400 images, then the network would not have sufficient incentive to learn to reject an interferer of that type, because complete misclassification of the RFI only leads to 0.1\% reduction in precision. A common method to reduce such false positives is fine-tuning, which refers to re-training of the network to a smaller dataset. However, fine-tuning is subject to over-fitting and can prove difficult in practice. We are developing a novel method to train our network to be more robust to such underrepresented RFI types. For this current work we simply manually reject all such false positives. 
 
In addition to RFI, background random noise presents another potential source of false detection. Even though distinction between RFI and FRB can be reduced with better training data, detection performance in background of noise is subject to trade off between recall and false alarm rate depending on threshold of detection. To test this explicitly, we ran our trained model on a 3-hour BL observation in X-band ($8-11.6$~GHz). The network returned only 7 positives due to occasional strong RFI. In other words, the observation, equivalent of around 120000 images, produced no false positive due to noise, thus indicating we are in the very low false positive regime in noise receiver operation characteristics (ROC).

\subsection{Inference Speed}
Inference speed is crucial in real time applications such as autonomous driving, where large number of images must be processed per second. For radio astronomy, inference speed is less of an issue, as we now explain. Because of the high noise in radio astronomy data, each pixel in the input spectrogram does not contain independent useful information. Therefore the first one or two convolutional layers should employ large convolutional kernels and large strides, which immediately reduces the data rate. In this analysis we fix the channel-reduction to a factor of 32. Our current model, without any inference acceleration techniques\footnote{Common techniques include pruning, which reduces number of neurons per layer, quantization, which represents the weights with lower precision, or compactification with \texttt{TensorRT}. } is capable of processing around 800 images per second on a Nvidia gtx1080 GPU, equivalent of around 70 seconds of observation.


In comparison, computations of HEIMDALL scales with number of DM trials. A search with 1200 DM trials process 3 seconds of observation per second. Thus in this setup our network inference appears 20 times faster than a brute force dedispersion. However, the different nature of the two algorithms makes it difficult to compare without ambiguity. A ``big O'' analysis is presented in \cite{connor18}. Though similar results applies to our network, actual considerations of hardware makes such analysis an oversimplification \footnote{Important considerations include number of memory read and write in addition to the amount of computation.}. We avoid further detailed comparisons of different algorithms in order to not distract from the main takeaway, that is neural network inference is likely more than fast enough for real time applications. 

\subsection{Pulse Verification} \label{sec:verification}

Despite their powerful capabilities, CNNs often suffer from lack of interpretability in decision making. With each detection, the neural network outputs a confidence. This confidence is learned with respect to the training set and should not be be assumed to translate directly onto real detections. Discrepancies between the simulation and real pulses can lead to biases in a neural network's predictions, and unfamiliar RFI are likely to activate the network as false positives. In this work we aim to conservatively verify detected pulses. Therefore we do not rely on the CNN to make the final decision on the reliability of a given detection. 

With a confidence threshold of 50\%, the network returns around 6000 candidates, roughly consistent with the 2\% false positive rate (out of $~200000$ samples) discussed in Section \ref{sec:finetune}. However, we find that all detections where one is able to identify a pulse by visual inspection\footnote{Success of visual inspection is subject to the resolution of the image. Here we inspect with 11MHz and 0.35~ms frequency and time resolution, respectively. } fall within neural network detection confidence of over 98\%. With a threshold of 98\% network confidence, and after manually rejecting obvious RFI false positives (see Section \ref{sec:model}), we arrive at around 300 candidates for dedispersion verification. 

Verifying band-limited pulses requires care. Signal to noise ratio (S/N) in frequency-integrated time series is often quoted as a measure of the reliability of a FRB detection. Before describing our verification procedure, we take a closer look at what these S/N actually show. There are many different S/N one can define. The commonly-employed frequency-integrated S/N over the full band is not useful here because for the weak and band-limited pulses the value can be lower than unity. A frequency-integrated S/N over sub-bands is not well-defined, due to its variability depending on the signal strength in the sub-bands chosen. A pixel-wise S/N over the two dimensional spectrogram is well-defined given certain frequency and time resolution. However, such a S/N only shows the confidence of the existence of a signal, and contains no information on whether the signal is a FRB or RFI. In fact, all of the above S/N measures only indicate the confidence that a signal exists. The dedispersion procedure builds a certain simple feature but still does not provide information on whether a detection is a quadratically dispersed FRB or, for example, a single-pixel RFI. Therefore a decision boundary based on S/N is inherently insufficient to distinguish RFI from FRBs. This is the theoretical reason that allows a CNN model to achieve better accuracy and lower false-positive rate than a S/N threshold-based dedispersion search, as it learns a better decision boundary from examples. 

As we explained in Section \ref{sec:finetune}, our network has a very low chance of mistaking noise for detection. Therefore to verify a candidate output of the CNN, we need to verify its morphology rather than its signal strength. The defining characteristic of a FRB burst lies in its quadratic dispersion. A faint but clearly quadratically dispersed signal is more likely to be a FRB pulse than a bright signal concentrated in frequency. In a blind search, we should verify the signal has expected spectral quadratic delay as a function of frequency. Here we have additional expectations of pulse DM being near the previously reported values between 553-569~pc~cm$^{-3}$ \citep{Spi14,Spi16,Sch16,Law17,vishal,Michilli:2018}. Thus as a final step to verify our detections, we check the signals have around the expected DM. 

\begin{figure}
\begin{center}
\includegraphics[width=0.5\linewidth]{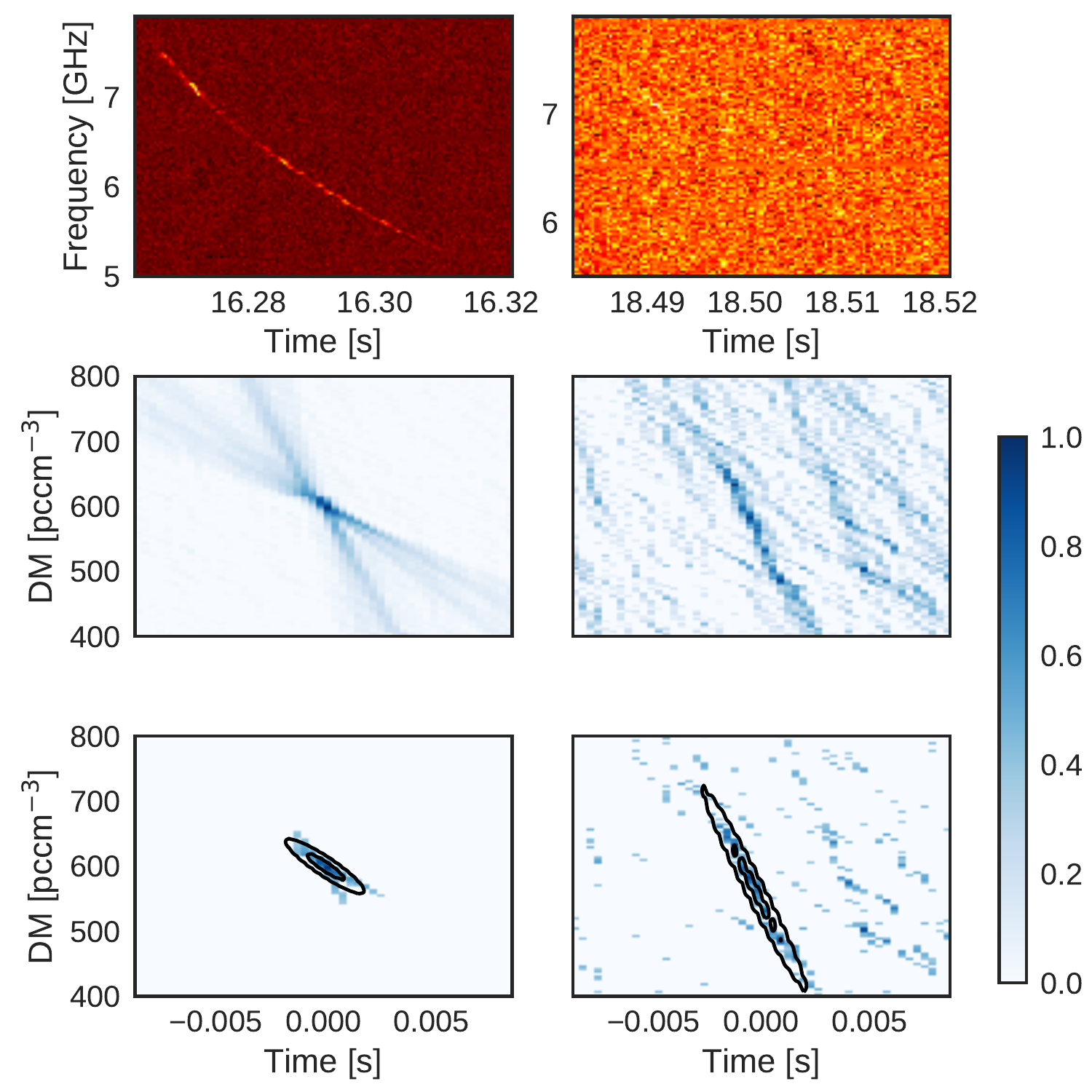}
\caption{Verification procedure for pulse 1 (left) and 2 (right). The first row shows the spectral data pre-dedispersion. Second row shows flux density as a function of DM and arrival time. Here both columns are normalized to their peak values, and for visual clarity only positive values are shown. In the third row a 2D Gaussian is fit to the pixels above $\sqrt{2}\sigma$ from peak to determine the DM and arrival time that maximizes S/N. }
\label{fig:verification}
\end{center}
\end{figure}

To verify the detections, we manually select the sub-bands where the candidate pulse is visible within the frame and perform a standard dedispersion search. We plot the frequency-integrated pixel intensity over DM and TOA and search for the DM that maximizes S/N, henceforth denoted as DM$_{S/N}$. While a smaller sub-band around the peak frequency of the pulse would lead to higher S/N, such S/N is less sensitive to variations in DM. Thus we choose sub-bands of 1.5~GHz at minimum to prioritize distinct fit of DM over high S/N value. For this analysis we search over DM of 400 up to 800 and fit a two dimensional Gaussian to thus determine the DM and arrival time that maximizes S/N for the given sub-band. If the Gaussian peak is observed between DM of 500~pc~cm$^{-3}$ and 700~pc~cm$^{-3}$ we keep the pulse\footnote{Complex spectral-temporal structure can render high power at a $DM_{SNR}$ that deviates from the more physical DM$_{\textrm{struct}}$ on the order of 50~pc~cm$^{-3}$. See \cite{vishal} for details.}. If no peak is seen at the expected DM, the candidate is discarded. An example of this procedure is illustrated in Fig. \ref{fig:verification} for two pulses. The two pulses both have peak frequency at 7.05~GHz. The one on the left, which is already reported in \cite{vishal}, has much higher S/N than the one on the right, which is a new candidate. In the second row we show the flux density as a function of DM and arrival times. Pulse 1 shows two linear wings in addition to the central peak due to its complex spectral structure. In the third row we fit two dimensional Gaussians to the pixels above $e^{-1}$ of the peak to determine the DM and TOA that maximizes S/N. The contours shown correspond to 1 and $2\sigma$ from center of the Gaussians. We see both pulses have a centroid fit within the acceptable bounds and are therefore confirmed detections. Sorting through the 300 candidates with this procedure produced 93 pulses.

To quantify the measure of confidence, we perform cross-validation of our procedure with other datasets. We apply the same detection and verification procedure to an additional set of Breakthrough Listen observation of 5 hours  at FRB\,121102 as well as 10 hours of observations of nearby stars, wherein no previous pulses have been detected. No signals from these observations survived our verification procedure. This shows that our procedure is conservative; it may discard some true detections but ensures the reliability of all pulses reported. 

\section{Pulse Properties} \label{sec:properties}
We summarize some of the main properties of detected pulses in Table \ref{tab:param}. Quoted MJDs are time of arrivals extrapolated to infinite frequency in a barycentric coordinate system. The TOAs are the corresponding values in seconds since the start of observation. DM$_{\textrm{SNR}}$ are the centroids of the Gaussian fits from Section \ref{sec:verification}. The fluence and flux density are calibrated with the radiometer equation and we shall detail the processes of their determination in the following section. The widths and sub-bands here are used to determine the fluences. In particular the widths are roughly the  full-width at 10\% max value determined in time series dedispersed to DM$_{\textrm{SNR}}$\footnote{In some cases, the value is difficult to determine algorithmically due to the noisy variations close to the peak. For these we manually increase the width to escape the local minima on the pulse profiles.}. Pictorially, these width are the distance between the vertical lines Fig. \ref{fig:butterfly}. These rough pulse widths are quoted mainly for reproducibility purposes and should not be confused with the intrinsic widths, which are full-width at half-max (FWHM) at $\textrm{DM}_{\textrm{struct}}$ \citep{vishal}. We do not quote the intrinsic widths here because for many of the weaker pulses the noise prohibits a reliable measurement. The peak frequencies are the locations of the channel (of width 11MHz) with the most power for each pulse. 
\begin{table*}
\tiny
\renewcommand{\arraystretch}{0.7}
\begin{threeparttable}
\begin{tabular}{lrrrrrrrr}
\hline
{} & MJD & TOA & DM$_{S/N}$ & Fluence & Flux Density& Subband & Width & $\nu_{peak}$ \\
Pulse & (57991+) & s & pc\,cm$^{-3}$ & Jy\,$\mu$s & mJy & GHz & ms & GHz \\
\hline \hline
1\tnote{*} & 0.4099040442 & 16.2278 & 600.1 & 606.1$\pm$21.2 & 763.5$\pm$10.5 & 4.1-8.0 & 1.43 & 7.02 \\
2 & 0.4099297208 & 18.4463 & 547.3 & 25.6$\pm$20.1 & 34.8$\pm$8.1 & 5.6-7.9 & 2.15 & 7.05 \\
3 & 0.4100218264 & 26.4042 & 646.1 & 50.5$\pm$16.8 & 53.9$\pm$8.3 & 5.1-7.5 & 1.43 & 7.02 \\
4 & 0.4100696130 & 30.5330 & 592.9 & 90.6$\pm$22.5 & 54.1$\pm$8.4 & 5.0-7.5 & 2.51 & 6.17 \\
5 & 0.4115584078 & 159.1649 & 615.3 & 53.0$\pm$19.0 & 45.7$\pm$8.4 & 5.0-7.5 & 1.79 & 5.56 \\
6 & 0.4120856083 & 204.7150 & 591.2 & 43.5$\pm$14.4 & 31.8$\pm$6.4 & 5.0-6.4 & 1.79 & 5.47 \\
7 & 0.4123121182 & 224.2854 & 607.0 & 26.4$\pm$14.6 & 44.9$\pm$8.3 & 5.1-7.5 & 1.08 & 7.02 \\
8\tnote{*} & 0.4127647200 & 263.3902 & 592.2 & 66.4$\pm$21.2 & 97.2$\pm$10.5 & 4.0-7.9 & 1.43 & 5.55 \\
9 & 0.4127648858 & 263.4046 & 631.4 & 67.3$\pm$20.6 & 56.9$\pm$9.1 & 4.0-6.9 & 1.79 & 5.45 \\
10 & 0.4129264759 & 277.3659 & 636.6 & 146.1$\pm$29.4 & 71.9$\pm$9.2 & 4.5-7.5 & 3.58 & 5.44 \\
11\tnote{*} & 0.4130198711 & 285.4353 & 570.7 & 56.6$\pm$10.3 & 138.0$\pm$7.2 & 6.1-7.9 & 0.72 & 7.03 \\
12 & 0.4133625264 & 315.0407 & 589.3 & 46.8$\pm$14.6 & 44.8$\pm$7.2 & 6.1-7.9 & 1.43 & 7.03 \\
13\tnote{*} & 0.4134587642 & 323.3557 & 585.3 & 363.2$\pm$18.6 & 452.1$\pm$10.6 & 4.0-8.0 & 1.08 & 7.06 \\
14\tnote{*} & 0.4137066537 & 344.7733 & 590.7 & 262.8$\pm$18.5 & 342.1$\pm$10.6 & 4.0-8.0 & 1.08 & 7.11 \\
15\tnote{*} & 0.4138370580 & 356.0402 & 589.4 & 264.6$\pm$21.5 & 318.2$\pm$10.6 & 4.0-8.0 & 1.43 & 7.03 \\
16 & 0.4161323109 & 554.3501 & 579.3 & 35.3$\pm$20.7 & 32.8$\pm$9.1 & 4.6-7.5 & 1.79 & 7.01 \\
17 & 0.4161852305 & 558.9223 & 527.0 & 46.2$\pm$18.8 & 37.5$\pm$8.3 & 5.1-7.5 & 1.79 & 7.02 \\
18 & 0.4162078619 & 560.8777 & 581.2 & 22.9$\pm$13.2 & 38.4$\pm$7.5 & 4.5-6.5 & 1.08 & 5.72 \\
19\tnote{*} & 0.4164367935 & 580.6574 & 557.9 & 36.2$\pm$18.5 & 68.9$\pm$10.6 & 4.0-7.9 & 1.08 & 5.47 \\
20\tnote{*} & 0.4166333622 & 597.6409 & 574.4 & 349.2$\pm$18.4 & 563.7$\pm$10.5 & 4.1-8.0 & 1.08 & 7.02 \\
21 & 0.4172693722 & 652.5922 & 578.0 & 52.6$\pm$14.8 & 73.2$\pm$8.4 & 5.0-7.5 & 1.08 & 5.80 \\
22 & 0.4173807378 & 662.2142 & 573.0 & 29.5$\pm$10.6 & 72.3$\pm$7.4 & 5.0-6.9 & 0.72 & 6.45 \\
23\tnote{*} & 0.4177147221 & 691.0704 & 559.7 & 35.2$\pm$14.7 & 54.2$\pm$8.4 & 5.0-7.5 & 1.08 & 5.75 \\
24\tnote{*} & 0.4178655533 & 704.1022 & 575.9 & 81.1$\pm$18.4 & 136.0$\pm$10.5 & 4.1-8.0 & 1.08 & 6.41 \\
25\tnote{*} & 0.4186272003 & 769.9085 & 579.9 & 97.8$\pm$21.3 & 128.8$\pm$10.5 & 4.1-8.0 & 1.43 & 7.61 \\
26 & 0.4190286984 & 804.5980 & 578.6 & 40.1$\pm$15.1 & 54.1$\pm$7.5 & 5.0-7.0 & 1.43 & 5.75 \\
27\tnote{*} & 0.4194498856 & 840.9885 & 607.8 & 44.6$\pm$13.9 & 50.8$\pm$6.9 & 5.0-6.7 & 1.43 & 5.75 \\
28 & 0.4194499461 & 840.9938 & 532.5 & 21.3$\pm$9.5 & 40.0$\pm$5.4 & 6.6-7.6 & 1.08 & 7.11 \\
29 & 0.4194499686 & 840.9957 & 594.5 & 36.2$\pm$12.0 & 49.6$\pm$6.8 & 5.1-6.7 & 1.08 & 6.22 \\
30 & 0.4194626399 & 842.0905 & 564.6 & 25.6$\pm$17.0 & 39.0$\pm$7.5 & 4.5-6.5 & 1.79 & 5.72 \\
31 & 0.4194628627 & 842.1098 & 546.2 & 29.5$\pm$15.0 & 29.3$\pm$7.4 & 4.6-6.5 & 1.43 & 6.23 \\
32 & 0.4208678576 & 963.5013 & 566.9 & 28.6$\pm$12.3 & 39.1$\pm$6.1 & 6.5-7.8 & 1.43 & 7.11 \\
33\tnote{*} & 0.4212129047 & 993.3134 & 566.2 & 60.7$\pm$15.1 & 126.8$\pm$10.6 & 4.0-8.0 & 0.72 & 7.11 \\
34\tnote{*} & 0.4217126673 & 1036.4929 & 565.7 & 78.5$\pm$15.1 & 185.8$\pm$10.6 & 4.0-7.9 & 0.72 & 5.67 \\
35 & 0.4221213510 & 1071.8031 & 595.8 & 68.2$\pm$17.0 & 64.0$\pm$7.5 & 4.5-6.5 & 1.79 & 5.72 \\
36 & 0.4221369638 & 1073.1521 & 596.0 & 39.7$\pm$18.7 & 55.7$\pm$8.2 & 5.5-7.9 & 1.79 & 6.90 \\
37\tnote{*} & 0.4229394568 & 1142.4875 & 582.4 & 131.2$\pm$23.8 & 214.7$\pm$10.5 & 4.1-8.0 & 1.79 & 6.20 \\
38 & 0.4230051411 & 1148.1626 & 579.3 & 26.3$\pm$26.3 & 35.5$\pm$10.6 & 4.0-8.0 & 2.15 & 5.77 \\
39\tnote{*} & 0.4242706565 & 1257.5031 & 629.3 & 53.0$\pm$16.8 & 45.6$\pm$7.4 & 4.6-6.5 & 1.79 & 6.23 \\
40 & 0.4245389367 & 1280.6826 & 570.0 & 17.4$\pm$9.0 & 31.2$\pm$5.1 & 4.6-5.5 & 1.08 & 5.25 \\
41 & 0.4259189384 & 1399.9147 & 653.7 & 28.7$\pm$12.7 & 24.5$\pm$5.1 & 5.1-6.0 & 2.15 & 5.75 \\
42 & 0.4263938242 & 1440.9448 & 622.9 & 67.5$\pm$18.3 & 56.1$\pm$7.4 & 5.1-7.0 & 2.15 & 6.17 \\
43\tnote{*} & 0.4265525154 & 1454.6557 & 563.6 & 156.5$\pm$15.1 & 392.2$\pm$10.5 & 4.1-8.0 & 0.72 & 6.17 \\
44 & 0.4283922705 & 1613.6106 & 618.4 & 74.6$\pm$22.0 & 57.4$\pm$8.2 & 5.5-7.9 & 2.51 & 7.14 \\
45 & 0.4285925362 & 1630.9135 & 569.7 & 36.2$\pm$18.4 & 63.0$\pm$10.5 & 4.0-7.9 & 1.08 & 4.78 \\
46\tnote{*} & 0.4304279044 & 1789.4894 & 564.1 & 51.0$\pm$18.5 & 70.4$\pm$10.6 & 4.0-7.9 & 1.08 & 5.67 \\
47 & 0.4310964736 & 1847.2537 & 596.7 & 57.4$\pm$19.3 & 27.1$\pm$6.4 & 5.5-6.9 & 3.23 & 6.16 \\
48 & 0.4316717532 & 1896.9579 & 584.0 & 41.2$\pm$16.9 & 42.4$\pm$7.5 & 5.0-7.0 & 1.79 & 6.08 \\
49\tnote{*} & 0.4319740079 & 1923.0727 & 572.0 & 162.9$\pm$18.5 & 208.4$\pm$10.5 & 4.0-7.9 & 1.08 & 5.66 \\
50 & 0.4322755018 & 1949.1218 & 588.0 & 48.0$\pm$15.1 & 58.0$\pm$7.5 & 5.0-7.0 & 1.43 & 5.52 \\
51 & 0.4348352197 & 2170.2814 & 646.8 & 24.3$\pm$16.7 & 28.8$\pm$7.4 & 5.5-7.4 & 1.79 & 5.73 \\
52 & 0.4351975268 & 2201.5847 & 617.3 & 56.1$\pm$14.9 & 61.4$\pm$7.3 & 5.6-7.5 & 1.43 & 7.09 \\
53 & 0.4365574396 & 2319.0812 & 578.0 & 56.4$\pm$12.8 & 71.5$\pm$7.3 & 6.0-7.9 & 1.08 & 7.12 \\
54 & 0.4378495339 & 2430.7181 & 580.0 & 24.4$\pm$14.4 & 41.0$\pm$8.2 & 5.5-7.9 & 1.08 & 6.91 \\
55 & 0.4384886807 & 2485.9404 & 569.0 & 22.4$\pm$11.3 & 42.0$\pm$6.4 & 5.0-6.5 & 1.08 & 5.52 \\
56\tnote{*} & 0.4393606779 & 2561.2810 & 593.8 & 234.2$\pm$23.9 & 236.2$\pm$10.6 & 4.0-8.0 & 1.79 & 7.11 \\
57 & 0.4462115554 & 3153.1968 & 578.4 & 18.0$\pm$9.1 & 38.7$\pm$6.4 & 5.5-6.9 & 0.72 & 6.16 \\
58 & 0.4469883544 & 3220.3122 & 544.7 & 13.1$\pm$9.3 & 39.5$\pm$6.5 & 5.5-7.0 & 0.72 & 6.44 \\
59 & 0.4479291105 & 3301.5936 & 582.9 & 52.3$\pm$24.4 & 33.4$\pm$9.1 & 4.5-7.4 & 2.51 & 5.58 \\
60 & 0.4480478257 & 3311.8506 & 593.5 & 25.8$\pm$12.8 & 17.9$\pm$5.1 & 4.5-5.4 & 2.15 & 4.97 \\
61\tnote{*} & 0.4484276509 & 3344.6675 & 580.2 & 137.5$\pm$18.4 & 210.0$\pm$10.5 & 4.0-7.9 & 1.08 & 5.58 \\
62 & 0.4501848147 & 3496.4864 & 580.1 & 19.7$\pm$9.0 & 46.3$\pm$6.3 & 4.6-6.0 & 0.72 & 5.39 \\
63 & 0.4504676214 & 3520.9209 & 581.3 & 34.3$\pm$16.4 & 40.5$\pm$8.1 & 5.6-7.9 & 1.43 & 7.47 \\
64 & 0.4530398739 & 3743.1635 & 588.9 & 67.2$\pm$16.8 & 84.3$\pm$8.3 & 4.5-6.9 & 1.43 & 4.78 \\
65 & 0.4585129735 & 4216.0393 & 572.6 & 21.8$\pm$14.4 & 29.0$\pm$6.4 & 4.6-6.0 & 1.79 & 5.53 \\
66 & 0.4608681160 & 4419.5236 & 581.8 & 27.2$\pm$11.0 & 54.0$\pm$6.3 & 6.5-7.9 & 1.08 & 7.53 \\
67 & 0.4637145928 & 4665.4592 & 588.5 & 47.1$\pm$13.0 & 73.1$\pm$7.4 & 5.3-7.2 & 1.08 & 6.75 \\
68 & 0.4639392978 & 4684.8737 & 629.7 & 43.9$\pm$15.2 & 38.8$\pm$7.5 & 4.5-6.5 & 1.43 & 5.48 \\
69 & 0.4639392697 & 4684.8713 & 629.2 & 36.7$\pm$16.1 & 57.3$\pm$9.2 & 4.5-7.5 & 1.08 & 5.48 \\
70 & 0.4701863590 & 5224.6198 & 698.1 & 20.5$\pm$9.1 & 45.2$\pm$6.4 & 5.1-6.5 & 0.72 & 6.22 \\
71 & 0.4711460870 & 5307.5403 & 528.8 & 27.2$\pm$16.1 & 30.8$\pm$6.5 & 5.5-7.0 & 2.15 & 6.62 \\
72 & 0.4728729109 & 5456.7379 & 589.9 & 20.1$\pm$11.5 & 25.5$\pm$5.1 & 4.6-5.5 & 1.79 & 5.06 \\
73 & 0.4800317945 & 6075.2655 & 581.0 & 35.3$\pm$16.9 & 46.9$\pm$7.5 & 4.5-6.5 & 1.79 & 4.83 \\
74 & 0.4886031426 & 6815.8299 & 653.1 & 26.5$\pm$14.5 & 22.3$\pm$5.1 & 6.0-6.9 & 2.87 & 6.42 \\
75 & 0.4931590534 & 7209.4606 & 600.0 & 18.5$\pm$16.6 & 26.6$\pm$7.3 & 4.5-6.4 & 1.79 & 5.11 \\
76 & 0.4975097257 & 7585.3587 & 600.0 & 50.0$\pm$15.1 & 68.1$\pm$7.5 & 4.5-6.5 & 1.43 & 4.83 \\
77 & 0.4983866613 & 7661.1260 & 574.2 & 24.6$\pm$12.7 & 26.8$\pm$5.1 & 5.1-6.0 & 2.15 & 5.94 \\
78 & 0.4988154972 & 7698.1774 & 600.0 & 51.5$\pm$18.7 & 41.7$\pm$8.3 & 4.5-6.9 & 1.79 & 5.53 \\
79 & 0.5108569357 & 8738.5577 & 648.4 & 24.1$\pm$10.7 & 55.5$\pm$7.5 & 5.5-7.5 & 0.72 & 7.33 \\
80 & 0.5131083628 & 8933.0810 & 581.4 & 96.4$\pm$13.1 & 128.4$\pm$7.5 & 4.5-6.5 & 1.08 & 4.83 \\
81 & 0.5192305588 & 9462.0387 & 582.1 & 49.8$\pm$21.5 & 110.6$\pm$7.5 & 5.5-7.5 & 2.87 & 6.91 \\
82 & 0.5192305898 & 9462.0414 & 576.2 & 65.9$\pm$15.2 & 98.8$\pm$7.5 & 5.5-7.5 & 1.43 & 6.91 \\
83 & 0.5593958795 & 12932.3224 & 609.0 & 29.7$\pm$15.1 & 76.5$\pm$10.5 & 4.1-8.0 & 0.72 & 6.98 \\
84 & 0.5628198735 & 13228.1555 & 575.7 & 58.5$\pm$15.7 & 102.2$\pm$9.0 & 5.1-7.9 & 1.08 & 6.83 \\
85 & 0.5694474415 & 13800.7774 & 569.4 & 69.6$\pm$21.4 & 44.8$\pm$7.5 & 4.5-6.5 & 2.87 & 5.95 \\
86 & 0.5710018046 & 13935.0743 & 581.5 & 49.2$\pm$14.9 & 43.4$\pm$7.4 & 4.6-6.5 & 1.43 & 5.67 \\
87 & 0.5744448464 & 14232.5532 & 622.9 & 17.7$\pm$13.6 & 18.3$\pm$5.1 & 6.0-6.9 & 2.51 & 6.42 \\
88 & 0.5881495443 & 15416.6390 & 575.2 & 47.8$\pm$13.1 & 67.9$\pm$7.5 & 4.5-6.5 & 1.08 & 5.58 \\
89 & 0.5905111113 & 15620.6784 & 595.5 & 42.1$\pm$14.6 & 64.4$\pm$8.3 & 5.1-7.5 & 1.08 & 6.41 \\
90 & 0.5921113250 & 15758.9369 & 579.2 & 51.8$\pm$13.1 & 86.2$\pm$7.5 & 5.0-7.0 & 1.08 & 5.94 \\
91 & 0.5954276026 & 16045.4633 & 564.7 & 20.9$\pm$9.3 & 59.0$\pm$6.5 & 5.5-7.0 & 0.72 & 6.25 \\
92 & 0.5957594875 & 16074.1381 & 556.8 & 29.2$\pm$13.1 & 32.9$\pm$6.5 & 5.5-7.0 & 1.43 & 6.25 \\
93 & 0.6022637323 & 16636.1049 & 560.6 & 21.3$\pm$8.7 & 60.1$\pm$6.1 & 4.5-5.8 & 0.72 & 4.92 \\

\hline
\end{tabular}
\begin{tablenotes}
  \item[*] Already reported in \cite{vishal}.
  \end{tablenotes}
\end{threeparttable}

\caption{Parameters for all pulse reported, including the barycentric coordinate time, corresponding TOA since start of observation, DM$_{SNR}$, fluence and flux density, the subband and width used to compute the fluences, and the peak frequency. The error estimate from fluence and flux densitities account for the expected noise for the given subband and width. Quoted fluence and flux densitities are determined from dedispersion to DM$_{SNR}$.  }
\label{tab:param}
\end{table*}

\subsection{Determination Procedures} \label{sec:fl}

In this section we describe the process with which we determine the subbands, flux density, fluence and time of arrivals quoted. 

First we describe the determination of flux density and fluence. Ideally, we would like to retrieve fluence from the entire 4~GHz of bandwidth at DM$_{\textrm{struct}}$. In practice this is impossible since the full-band S/N for many band-limited pulses can be lower than unity. Since fluence is conserved over DM we dedisperse to DM$_{\textrm{SNR}}$ for a better determination of fluence\footnote{Note although the noise induced S/N is the same for either DM choice, lower S/N leads also to more uncertainty in the width, therefore we use the DM$_{S/N}$ for a more reliable fluence determination.}. In order to retrieve the maximum fluence we search over ranges of frequency to find the widest sub-band around the peak frequency with a minimum S/N of 3. After the appropriate sub-bands are chosen, fluence can thus be estimated with corresponding noise-induced errors. The flux densities and widths quoted here are also from DM$_{\textrm{SNR}}$.  

To determine the arrival times, we dedisperse to DM$_{\textrm{struct}}$ of 565~pc~cm$^{-3}$ with the sub-bands determined above and record the time stamps of the energy peak. Due to the presence of multiple sub-pulses, this procedure records the arrival times of the strongest sub-pulse, and thus introduces a minimum uncertainty of around 2~ms, based on previously reported intrinsic pulse widths \citep{vishal}. 

\section{Discussion} \label{sec:discussion}

\subsection{Pulse Structure}
Extended figures on all pulses presented in this paper are located in App. \ref{sec:FigEx}. Many pulses of FRB\,121102 from this observation exhibit complex structures in both frequency and time, as already reported in \cite{vishal}. In that work, reported substructures within a pulse extend over width of up to $\sim2$~ms. The tilt of the wings in Fig. \ref{fig:butterfly} corresponds to the frequency range of the pulses, where higher frequencies corresponds to more vertical tilts \footnote{This correspondence has purely geometric explanations, hence we will not delve into the details here. }. For example, in pulse 1 we see three sub-structures at three distinct frequencies while in pulse 2 we only see one. Furthermore, Fig. \ref{fig:butterfly} shows that for complex pulses such as pulse 1, 13, 14 and others, there exits a DM$_{S/N}$ that stacks the multiple components into one. This is possible only because the different components are close in time and monotonically increase in peak frequency. This suggests that these cases are different sub-images of the same pulse, possibly caused by propagation effects such as lensing. As \cite{vishal} explains, the S/N-maximizing DM, which stacks all sub-pulses together, may not be indicative of the physical DM characteristic of the dispersion process. Instead, an approximate structure DM (DM$_{\textrm{struct}}$), taken as the average values of all DM$_{\textrm{S/N}}$, separates the sub-pulses and can be regarded as a closer approximation of the ``physical'' DM. For the rest of this paper we use the value of DM$_{\textrm{struct}} = 565$~pc~cm$^{-3}$ as determined by \cite{vishal}. 

A few of the new pulses in this work are clustered up to 10~ms time scale. It is unclear whether these multiplets are different pulses or substructures of the same pulse. Unlike substructures already discussed in \cite{vishal}, these sets of closely timed signals, i.e. \{8,9\}, \{27,28,29\}, \{30, 31\}, \{81, 82\}, do not have a unifying DM that stacks the sub-pulses. The pair \{81, 82\} is particularly interesting due to their close proximity and yet lack of an unifying DM. The two pulses clearly have power in overlapping frequency channels. If they are indeed independent pulses, they would be by far the closest pair of pulses seen from FRB\,121102, with a TOA separation of 2.56~ms. However, it is also possible that they are sub-images created by a different physical process than pulse 1. Other multiplets, i.e. \{8,9\}, \{27,28,29\}, have inter-spacings on order of 10~ms. If they are sub-pulses, this time scale indicates a lower bound on the uncertainty of time of arrivals. If they are different pulses, the time scale roughly indicates an upper bound on any possible periodicity. In addition, the triplet also shows components ranging from low frequency to high frequency, and back to low frequency, the first time such a structure has been observed.

\subsection{Parameter Statistics} \label{sec:stats}

\begin{figure}
\includegraphics[width=\linewidth]{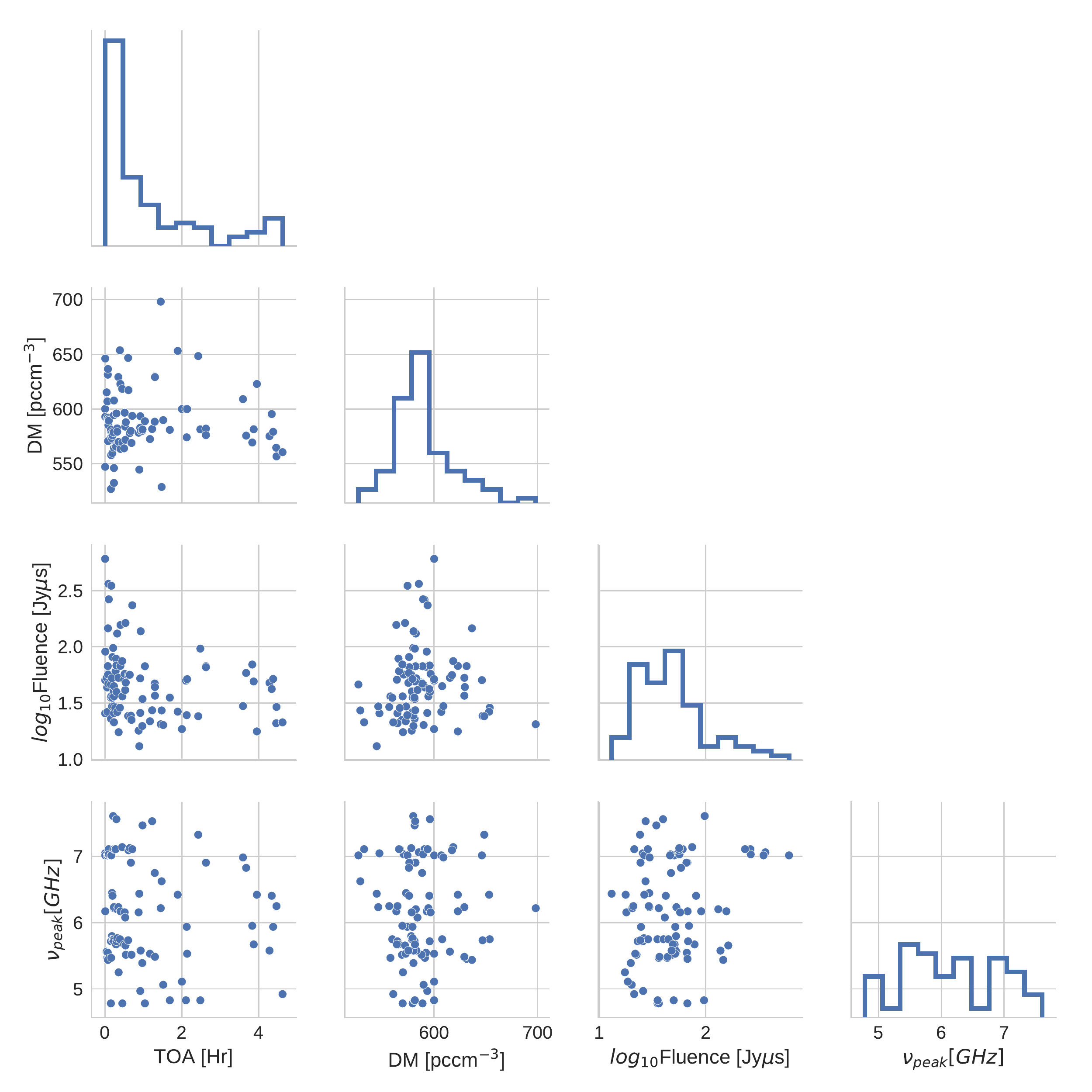}
\caption{Pairwise comparisons of arrival time, DM$_{S/N}$, fluence, and the peak frequency. Diagonals are histograms of the quantities. Each off-diagonal scatter plot relates a pair of quantities for all pulses. }
\label{fig:pairplot}
\end{figure}

In Fig. \ref{fig:pairplot} we show scatter plots of TOA, DM$_{\textrm{SNR}}$, fluence, and peak frequency. On the diagonal we show histograms of each quantity and the off diagonal shows the two relating quantities of each burst. The DM$_{\textrm{SNR}}$ appear scattered around an average of 575~pc~cm$^{-3}$. We still use the values from \cite{vishal} as DM$_{\textrm{struct}}$ since the scatter for lower energy pulses are large. 

\subsubsection{Peak Frequency}
In \cite{vishal}, the authors point out that the frequency distribution of the burst energy appears to be multi-modal. Here we see the same behavior, most notably from the $\nu_{peak}$ vs. TOA scatter plot in Fig.\ref{fig:pairplot}, where the pulses from the first hour show three to five distinct clusters.  Expectation-maximization (EM) fit with a 5-component Gaussian Mixture Model (GMM) shows these clusters are centered at 7.54~GHz, 7.05~GHz, 6.23~GHz, 5.62~GHz, and 4.91~GHz, respectively. To see if this behavior is caused by instrumental effects, we compare with an observation of pulsar J0332+5434 in the same session. The pulsar pulses extend over the full band and do not exhibit similar modulations. Comparison with coherently dedispersed dynamic spectra of the strong pulses \citep{vishal}, we see that these frequencies correspond with the peak frequencies of the sub-pulses, and are thus a type of characteristic frequencies of the process that created them. It is unclear if these characteristic frequencies persists for longer than the first hour. Follow-up observations during a similar active phase of FRB\,121102 will help to provide an answer. 

In addition, we note that the behavior of the peak frequency possibly varies with time in a non-stochastic manner. The first three pulses are spaced 8 seconds apart and all peak around 7~GHz in frequency. It is possible that a certain time scale exists that characterizes the variation of pulse frequency structure with time. However, more data are needed to further constrain the time scale of variability, if any. 

\subsubsection{Fluence}

The fluence vs. TOA plot shows a higher density of pulses towards the beginning of the observation for all fluences.  High pulse rate is thus coincidental with the high energy of the pulses detected. The cause of this behavior is subject to interpretation. It is possible that the intrinsic fluence distribution is constant with time, and the excess of bursts at the beginning of observation samples high energy bursts from the tail of the distribution. It is also possible that an occasional high energy burst is followed by many weaker ``after bursts". 

The fluence histogram shows fall-off of detections on the low energy end. It is unclear whether the fall off is intrinsic or due to detection bias. If the fall off is detection bias, and the intrinsic distribution of pulses is dominated by the low-energy end, then large number of low energy pulses must be undetected from the later four hours of observation. 

The fluence vs. $\nu_{peak}$ plot shows over an order of magnitude variation of pulse flux density among pulses with similar frequency structure. If propagation effects leave distinct frequency-imprints, such that the same amount of amplification applies to pulses with the same frequency structure, then this variation in flux density is likely due to the source itself. Though we point out that this assumption about propagation models has not yet been confirmed. 

\subsection{Pulse Rate} \label{sec:TOA}

In this section, we quantify the rate of detection. If each occurrence of an event is independent and random with a fixed rate $r$, the interval $\delta$ between each consecutive occurrence follows stationary Poissonian statistics:
\begin{equation}
P(\delta|r)=re^{-r\delta}.
\end{equation}
The occurrence of giant pulses from the Crab nebula has been observed to be locally consistent with Poissonian statistics \citep{ramesh10}, with a rate varying smoothly on daily timescales \citep{Lundgren95}. The phase-duration of nulling and state-switching pulsars have also been shown to follow stationary Poisson process \citep{vishal12, cordes13}. In comparison, the observed pulses from FRB\,121102 to date have been clearly non-Poissonian on timescales spanning multiple observation sessions \citep{Pen17}. The non-Poissonian nature of detections could be either intrinsic or detection bias. 

\begin{figure}
\begin{center}
\includegraphics[width=0.7\textwidth]{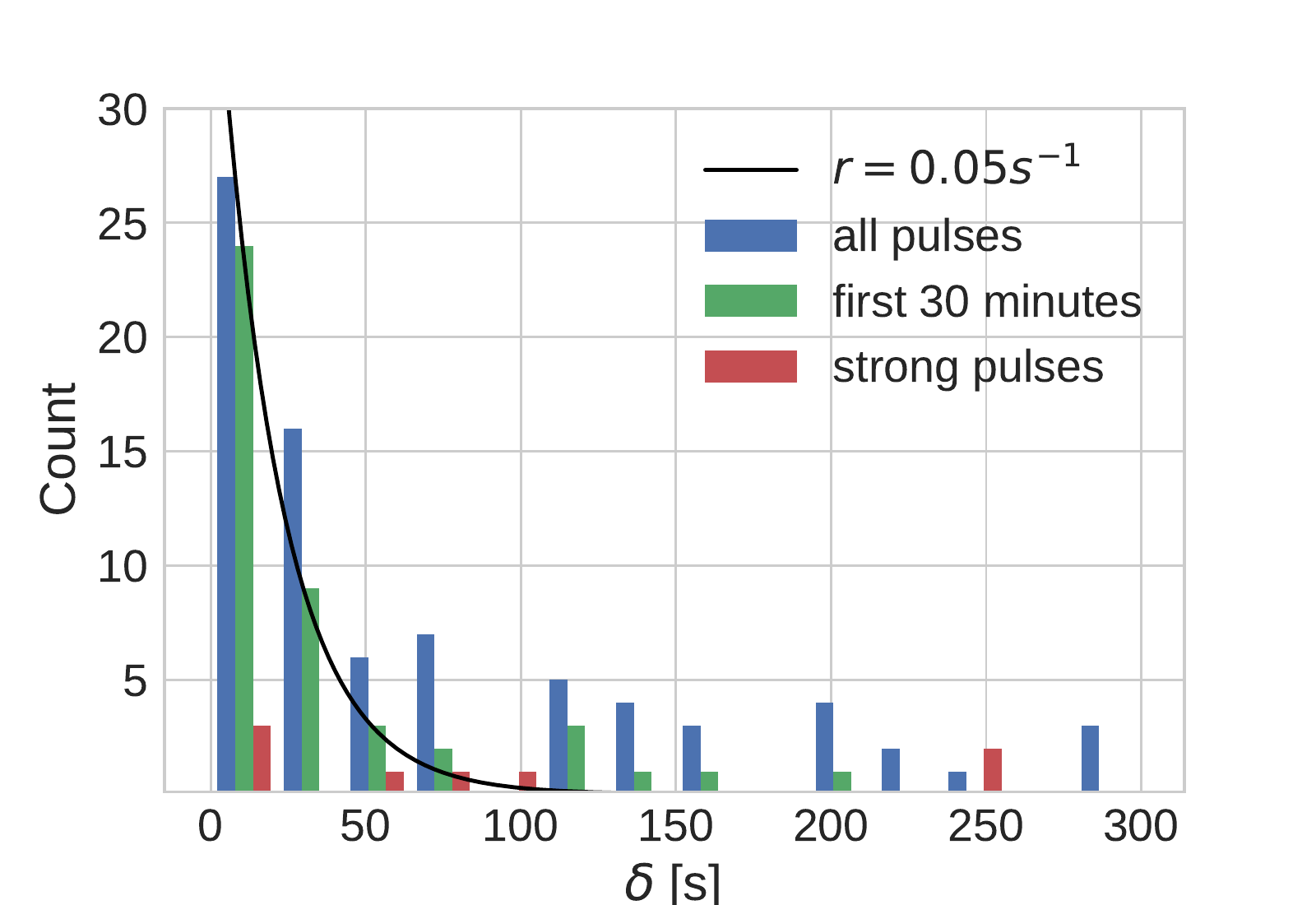}
\caption{Distribution of intervals between consecutive detections, compared against Poissonian expectations. Shown are histograms from all pulses (blue), pulses from the first 30 minutes (red), and the 15 pulses with the highest fluence. The bars are shown side-by-side for visual clarity, while the actual bins completely over lap, with constant bin widths of 20 seconds for all the histograms. Only intervals less than 300 seconds are shown. The skew towards the tail end indicates clustering of pulses missing from detection. }
\label{fig:poisson}
\end{center}
\end{figure}

In Fig. \ref{fig:poisson} we show the distribution of the intervals in our observation. Three histograms are shown, one for all the pulses, one for the 45 pulses from the first 30 minutes of observation, and one for the 15 pulses with the highest fluence. The bars are shown side-by-side for visual clarity. The actual bins completely overlap with constant bin-widths of 20 seconds for all the histograms. Also shown for comparison is a Poissonian expectation with $r=0.05\,s^{-1}$. 

From Fig. \ref{fig:pairplot} we already see that the rate of detection is not stationary, with almost half of the detections being from the first 30 minutes of observation. Nevertheless, on the timescale of a five hour observation the observed intervals are more consistent with Poisson statistics than previously reported. This can also be seen in the direct comparison with the distribution for the 15 strongest pulses, for which the resemblance to Poissonian distribution is much harder to recognize. Thus observational bias likely played a major role in previously reported behavior. 

The observed distribution even for the first 30 minutes still deviates from Poissonian expectations by a skew towards a longer tail. Even though we have the largest number of detections in a single observation of FRB\,121102 to date, we must assume that an unknown number of pulses are still missing from detection. The long tail can be explained by the intuitive expectation that the missing pulses are clustered rather than randomly scattered. Many scenarios can account for the clustering of missing pulses, including observational bias caused by non-stochastic variations of the pulse energy and pulse frequency structure, such that the pulses were outside the band of observation or too weak to be detected, in addition to the possibility of no emission. 

\subsection{Periodicity} \label{sec:period}

Due to its unique nature as a repeating FRB, FRB\,121102 has generated much interest in searching for periodicity of its pulses \citep{Spi16,Katz2017}.  The
measurement of an intrinsic emission period, if it exists, would be highly suggestive of theoretical models involving a rotating source (e.g. \citealt{Met17,Cor16a,KatzReview,Lyutikov16,popov18,Kulkarni15,Yang2016}). Lack of periodicity
does not necessarily exclude rotating models, but at the same time permits models that predict intrinsically non-periodic emissions (e.g. \citealt{Katzfunnels,Huang2016}), and those predicting periods much smaller than observation sensitivity (e.g. \citealt{axion}).  All previous analysis have reported non-detection
of significant periods (e.g. \citealt{Spi14,Spi16,Sch16}). We have performed a Fourier-domain acceleration
(\texttt{-zmax 300}) and jerk (\texttt{-wmax 1200}) search using a harmonic
summing of 16 with \texttt{PRESTO} \citep{2002AJ....124.1788R} and a time-domain
acceleration search with
\texttt{riptide}\footnote{https://bitbucket.org/vmorello/riptide} over a range
of dedispered time series ($560-570$~pc~cm$^{-3}$). No significant candidates
were found in either of these methods. 

Furthermore, with the highest rate of pulses to-date, we have an enhanced opportunity to quantify the statistical significance of detection or non-detection. More precisely, we can constrain the likelihood of aperiodicity, and for any notable periods, constrain the significance of detection. For the rest of this analysis, we focus on periodicity searches based on the pulse time of arrivals (TOA). The advantage in working with time stamps of arrivals, instead of spectral or voltage data, is that the confidence of any statistical tests can be obtained through unambiguous simulations. 

There are two main challenges for any period search based on the
time of arrivals (TOA). The first is the uncertainty in the time stamps $\sigma_t$. As discussed in Section \ref{sec:TOA}, the apparent TOA have measurement error of around 2~ms due to intrinsic width of the pulses. In
addition to measurement error, many physical scenarios can perturb the time stamps, smearing out periodicity in the time of emission (TOE). These include propagation effects such as lensing, or acceleration effects at
the source. We do not confine ourselves to specific models at this stage and
simply model the effect of all perturbations in the effective uncertainty $\sigma_t$. Another main challenge is the (unknown) number of missing pulses.
The observed pulses vary greatly in energy as well as frequency structure. 
This means that in search of an intrinsic period, one must assume that there are
an unknown number of unobserved pulses, where again potential causes include pulse energy below
detection threshold, or pulse peak frequency outside detection bandwidth. Since the total number of detection is fixed, the number of missing pulses translates directly to the value of the candidate period $t_p$. As we shall see, the ratio of these two quantities $\alpha=t_p/\sigma_t$ is an important parameter in period detection. 

\subsubsection{Hypotheses of Periodicity}
We construct two hypotheses:
\begin{itemize}
\item $H_0$: The times of arrival are not periodic with any period.  
\item $H_1$: The times of arrival are periodic with some period $t_p$ and uncertainty $\sigma_t$. 
\end{itemize}
To quantify the significance of a candidate period, we are interested in the confidence of rejecting $H_0$. To address the confidence of aperiodicity, we are interested in the confidence with which we can reject $H_1$. 

There are many ways to search for periodicity in times stamps, including Fourier transform, autocorrelations, and histograms of separations\footnote{We display such histograms in Fig. \ref{fig:poisson}, but here we are interested in behaviors on much shorter time scales. }. However, the sensitivity of these methods tend to be poor when a large portion of the events are missing from detection. Here we follow a procedure similar to the one in \cite{eperiod}, where the authors show that their method outperforms the aforementioned ones in a wide range of cases where the observation is incomplete.

To proceed, we fold the TOAs with a series of trial periods. If $t_p$ is a period that fits the data well, the folded pulse phases would show a strong unimodal deviation from uniformity. We construct the specific hypotheses for a given trial period:
\begin{itemize}
\item $h_0(t_p)$: The time of arrival are not periodic with $t_p$. The folded pulse phases are not unimodally distributed. 
\item $h_1(t_p)$: The time of arrival are periodic with true period $t_p$ and uncertainty $\sigma_t$. The folded pulse phases are unimodally distributed.
\end{itemize}


One cannot test an infinite number of trial periods. The list of trial periods should cover the range such that the closest period to the true period has an error that, when propagated to the end of observation ($T_{obs}$) is less than the assumed uncertainty in TOA ($\sigma_t$):
\begin{equation}
\delta t_p \cdot \frac{T_{obs}}{t_p} < \sigma_t. 
\end{equation}
Solving the differential equation $\frac{dt_p}{t_p}=\frac{\sigma_t}{T_{obs}}dn_t$, we get the total number of trial periods over a search range of $[t_{p_1}, t_{p_2}]$ is:

\begin{equation}
n_t = \frac{T_{obs}}{\sigma_t} \ln \frac{t_{p_2}}{t_{p_1}}.
\end{equation}
In theory, since the relative resolution $\delta t_p/t_p$ is fixed, we only need to search from $t_{p_1}$ to $2t_{p_1}$ to cover all periods greater than $t_{p_1}$. This is because any period greater than $2t_{p_1}$ is an integer multiple of a period in this range. 
In practice, however, the measurement uncertainties may smear out unimodal behavior at small trial periods, therefore we test trial periods up to order of a second, close to the separation between the first two pulses. 
It is important to note that not all of these periods are independent, meaning periods that are close to each other lead to correlated measure of validity (to be defined in Section \ref{sec:rayleigh}). The search is exhaustive so long as the trial period separations are smaller than the characteristic correlation length. 



\subsection{Rayleigh's Test} \label{sec:rayleigh}

The rest of our technique differs from \cite{eperiod}. There exists many  measures to detect non-uniformity on the unit circle (see for
example \citealt{mar1999} Chapter 6 and \citealt{Pew2013}). Here we would like to test the widest range of possible periods. 
In order to be sensitive to periods on millisecond time scales, we need to account for extreme scenarios
where less than 0.01\% of the events are detected. We therefore adopt the {\it mean resultant radius}. While the probability measure of
\cite{eperiod} is applicable for a wide range of distributions, the mean
resultant radius is more powerful to test unimodal departure. Its typical
usage, known as Rayleigh's test, has been proven to be the most powerful test
when the alternative is von Mises type distributions \citep{wandw}.
The mean resultant radius $\bar{R}$ is defined as follows:
\begin{equation}
\bar{R} = \sqrt{a^2+b^2}, 
\end{equation}
where $a$ and $b$ are the first angular moments of the pulse phases:

\begin{align}
a & = \frac{1}{n}\sum_i\cos\theta_i, \\
b & = \frac{1}{n}\sum_i\sin\theta_i, 
\end{align}
where $\theta_i=\theta_i(t_p)$ is the folded pulse phase of the ith pulse and $n$ is the number of pulses. As the name suggests, $\bar{R}$ is the length of the mean of a series of unit vectors. Thus defined, $\bar{R}$ takes value between 0 and 1, and takes 1 if and only if all pulse phases are aligned. If the underlying phases $\theta_i$ are uniformly distributed, it can be shown that $\bar{R}$ follows the Rayleigh distribution:
\begin{equation}
p(R, n) = 2nR e^{-nR^2}.
\label{eq:Rayleigh}
\end{equation}
For the rest of this analysis we denote the mean resultant radius from the detected TOAs with $\bar{R}$, and those from hypotheses with $R$. The next two sections test each of the two hypotheses as null hypothesis. We will, in each case, construct the relevant statistical $p$-values, which in this case is the probability $P(R>\bar{R}|H)$. 

\subsubsection{Top Scoring Periods}

Note strictly speaking $h_0$ does not imply that the pulse phases are uniformly distributed; if $t_p$ is a close rational fraction of the true period, multi-modal distribution can be observed. However, we are interested in detecting unimodal distribution only. By searching through an exhaustive list of periods we aim directly test the true period, if it exists. Thus for simplicity we can ignore this distinction and treat $h_0$ as if uniform distribution of pulse phases is implied.

We compute the $\bar{R}$ values of trial periods from 2 to 1000~ms. It may be tempting to compare these values with the Rayleigh distribution. However, Rayleigh distribution applies for a single hypothesis. The multiple trial periods require multi-testing correction. Since the precise number of independent periods is hard to obtain, we avoid the need for such corrections by simulations. We simulate 1000 incidents of 93 random time of arrivals for comparison. We split the periods into logarithmic intervals of $[2^{i}, 2^{i+1}]$ milliseconds, with i ranging from 1 to 9. For each of the 1000 simulations we perform the same period search to get the distribution of the highest scores for each logarithmic interval. The distribution of time of arrivals is highly non-uniform, with most pulses discovered towards the beginning of the observation. Thus to account for this bias we sample our random trials from the empirical distribution of time of arrivals with 20 bins. 
We quote the top periods for each logarithmic interval of trial period in Table \ref{tab:top_scores}. Comparisons with the empirical cumulative distributions from simulations allows us to quote the $p$-value $P(R>\bar{R}|H_0)$, or confidence of rejection $P(R<\bar{R}|H_0)$. 

The fact that the 8th of the top scoring periods rejects $H_0$ by 97.5\% should not be over-interpreted. To see the overall confidence that such as period rejects $H_0$, we again need to account for multiple testing. Suppose the 9 different $\bar{R}$ are independently sampled from distributions consistent with $H_0$, we ask the question:``What is the probability that none of the top results reject $H_0$ by 97.5\%?". The answer is $0.975^9\approx0.79$. Thus, assuming independence of the 9 periods, the confidence with which our best scoring period rejects $H_0$ is only 79\% ($p$-value 0.21). 
\begin{table*}
\begin{center}
\begin{tabular}{c | c | c |c }
\hline
$\log_2$ & $t_p [ms]$ & $\bar{R}$ & $P(R<\bar{R}|H_0)$\\ 
\hline
1 &    2.169175 &  0.341 & 0.817\\
2 &    5.127793 &  0.326 & 0.801\\
3 &   10.141185 &  0.308 & 0.723\\
4 &   17.622554 &  0.321 & 0.925\\
5 &   48.142379 &  0.291 & 0.814\\
6 &   72.433702 &  0.230 & 0.235\\
7 &  132.477494 &  0.263 & 0.823\\
8 &  276.327420 &  0.300 & 0.975\\
9 &  568.577720 &  0.170 & 0.176\\
\hline
\end{tabular}
\caption{Top scoring periods from 2 milliseconds to 1 second. The $i$th row is the top scoring period in the range  $[2^{i}, 2^{i+1}]$(milliseconds). The second column shows the candidate period in milliseconds, the third column shows the mean resultant radius, and the last column shows the confidence with which the period rejects $H_0$ (before correcting for multiple testing). }
\label{tab:top_scores}
\end{center}
\end{table*}

\subsubsection{Period Exclusion}
\begin{figure}
\begin{center}
\includegraphics[width=0.8\linewidth]{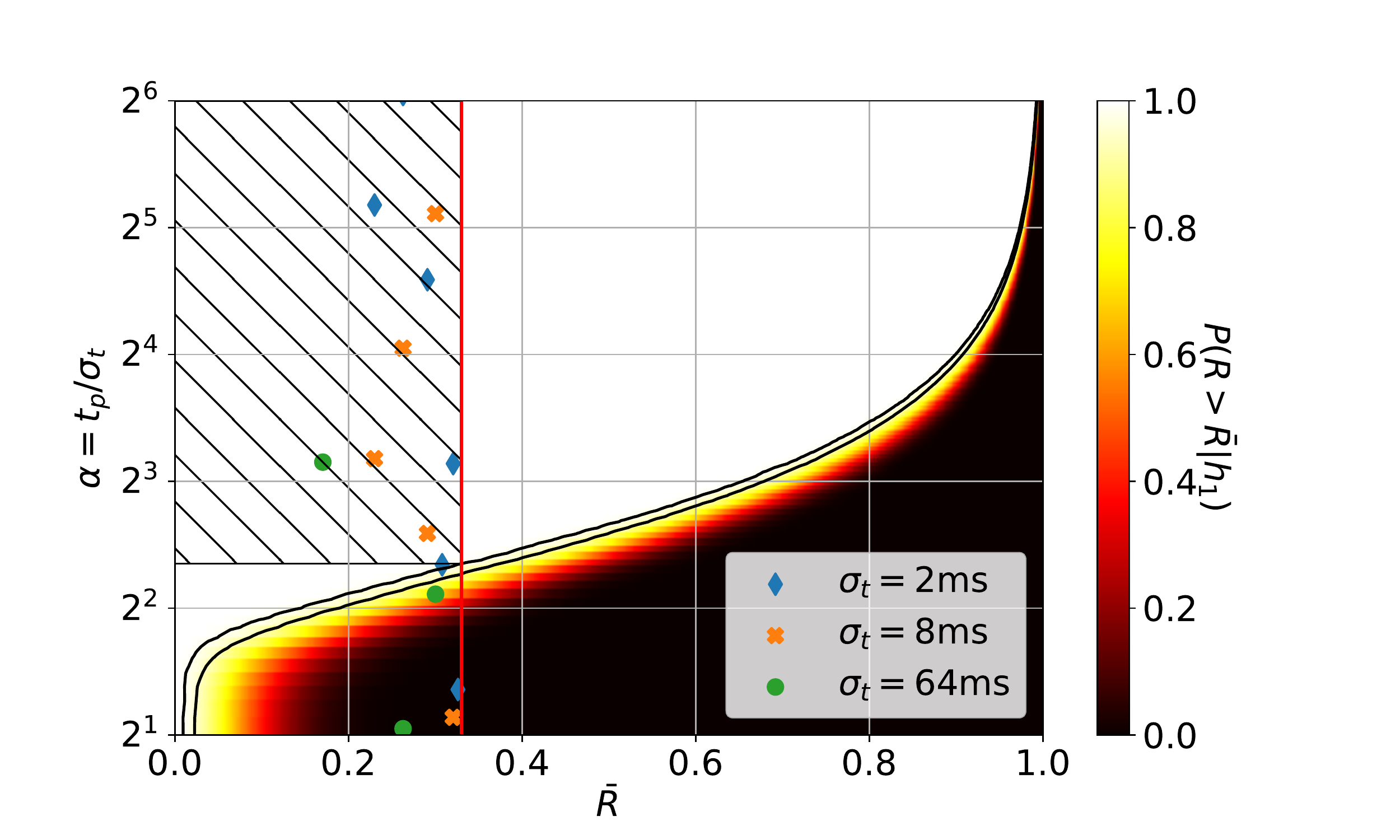}
\caption{Cumulative distribution of $R$ for 93 pulse phases given that $t_p$ is a true period, as a function of the parameter $\alpha=t_p/\sigma_t$.  Larger $\alpha$ means less perturbation in the TOAs, thus a distribution towards higher $R$. The two black curves trace out values of 0.95 and 0.99. Any best-fit period that lies above the black curves can be excluded with 99\% confidence. Scatter points show best fit periods from Table \ref{tab:top_scores}. The vertical line shows $R=0.33$, a rough upper bound of all $R$ from the observed TOAs. At its intersection with the confidence contours, we exclude any periods larger than $5.1\sigma_t$ with 99\% confidence, as indicated by the shaded region.}
\label{fig:h1}
\end{center}
\end{figure}

In this section we quantify our sensitivity to the top scoring periods. We compare observations with expectation from $H_1$ to see if a period exists, under what conditions we should expect to detect it with significance. To keep our test simple and model-agnostic, we introduce the only unknown parameter, i.e. the time stamp uncertainty $\sigma_t$. The distribution $P(R|h_1(t_p))$ is a function of the dimensionless parameter $\alpha=t_p/\sigma_t$. Fig. \ref{fig:h1} shows the cumulative distribution for $h_1$ and $n=93$ pulses. The black contour traces out $P(R>\bar{R}|h_1)=95\%$ and $99\%$, respectively. For different $\sigma_t$, the top scoring periods in Table \ref{tab:top_scores} slides along the vertical axis. While as we mentioned the measurement uncertainty for the TOAs are around 2~ms as discussed, perturbation caused by propagation effects can be much larger. For three different assumed uncertainties $\sigma_t$, we show the best scoring periods in the range $2\sigma_t<t_p<2^6\sigma_t$. If a period lies above the black contour, $h_1$ implies over 99\% probability that the score is lower than expected from $H_1$. The red line is an approximate upper bound of $\bar{R}=0.33$ for all the candidate periods. The intersection of the vertical line with the confidence contour marks the region of periods that we can exclude with 99\% confidence, as indicated by the shaded rectangle: $t_p>5.1\sigma_t$. With more pulses, this contour would shift to lower $\alpha$, allowing for stricter limits on periodicity exclusion.

We point out that the confidence with which we exclude each period is different from the confidence of aperiodicity. For the latter we need to correct for multiple testing. The probability of aperiodicity is:
\begin{equation}
P_a = \prod_{t_p} P(R>\bar{R}|h_1(t_p)),
\end{equation}
where the product is over all independent periods in the range of interest. The exact number of independent periods is tricky to determine. However, from the proximity between the 95\% and 99\% contours, we see the error rate drops off quickly. Since the shaded region of exclusion borders the contour only on one corner, we can expect the correction to be small. 

\subsubsection{Model Dependence}
If we take $\sigma_t$ to be the measurement uncertainty in TOAs, then our constraint applies to the periodicity of apparent TOAs. With a lenient assumption of 3~ms uncertainty, we can exclude periods larger than 15~ms in the barycentric arrival times. Some candidate periods reject both $H_1$ and $H_0$ with high confidence for the measurement uncertainty of 2~ms. While the confidence of rejecting $H_0$ is only 79\% after multiple test correction, there remains an alternative explanation where the intrinsic uncertainty is much larger. With given physical model, we can also derive model-dependent constraints on intrinsic periodicity. Many physical scenarios potentially smear out quasi-periodic time stamps, such as acceleration of source or lensing during the propagation. Constraining specific model parameters is beyond the scope of this paper. Here we give a brief example in the simplest scenario.  

Acceleration has been extensively observed in pulsars. A constant acceleration $a$ perturbs the period over the time of observation by $\sigma_a\sim\frac{aT_{obs}}{c}t_p$. For time dependent perturbation, restricting to the 45 pulses from the first 30 minutes of observation leads to exclusion of $t_p>6\sigma_t$. Thus we can exclude accelerating periodic effect if the acceleration $a<\frac{c\sigma_t}{t_pT_{obs}}=1.5\times10^5\frac{\sigma_t}{t_p}$~ms$^{-2}<2.5\times10^4$~ms$^{-2}$.



\section{Conclusion} \label{sec:conclusion}
Modern machine learning ushers in a new era of sensitive detection of fast radio transients. In this paper, we demonstrate the first application of a neural network for direct detection of fast radio bursts in spectral-temporal data. The approach shows potential advantage in terms of sensitivity and computational speed over dedispersion pipelines. The unprecedented abundance of detections from a single observation allows for explorations in trends of pulse fluence, pulse rate, and pulse frequency structure. We report the first detection of multiplets of pulses within spans of 10~ms to 20~ms that show non-monotonic variations in frequency structure. We pay special attention to the search for periodicity and introduce a new method for detecting periods in time of arrivals when most of the pulses are potentially unobserved. Our method allows us to quantify the significance of null detection and exclude with 99\% confidence all periods greater than 10~ms in the time of arrivals when the time stamp measurement uncertainty is 2~ms. 

Deep learning is a very active field of research that has found successful application across the board in academia and industry. We believe deep learning methods have the potential to completely surpass traditional algorithms, and even humans, for reliably identifying radio transients, as well as other similar signal detection tasks such as those occurring in gravitational wave astronomy \citep{gwave} and the search for extraterrestrial intelligence. There remains many caveats for applying a deep learning technique for direct real time detection in a general survey. Developing such a general pipeline and comparing its performance to other existing pipelines is outside the scope of this work.



We attribute the abundance of detections in this analysis to a combination of high sensitivity detection and wide bandwidth observation. The abundance of pulses and high spectral and energetic variations in this analysis suggests that previous surveys may have underestimated both the abundance of FRBs and the percentage of repeaters. Similar techniques applied to archival data and new broadband follow-ups could soon uncover an unforeseen abundance of both repeating and non-repeating sources,  accelerating the path to solving the mystery of FRBs. 

\section*{Acknowledgements}
Breakthrough Listen is funded by the Breakthrough Initiatives (\href{http://breakthroughinitiatives.org}{breakthroughinitiatives.org}). Y. Zhang thanks Dan Werthimer, Deepthi Gorthi, and Liam Conner for helpful discussions. VG would like to acknowledge NSF grant 1407804 and the Marilyn and Watson Alberts SETI Chair funds.

\appendix
\section{Extended Figures \label{sec:FigEx}}
\begin{figure}
\includegraphics[width=0.9\linewidth]{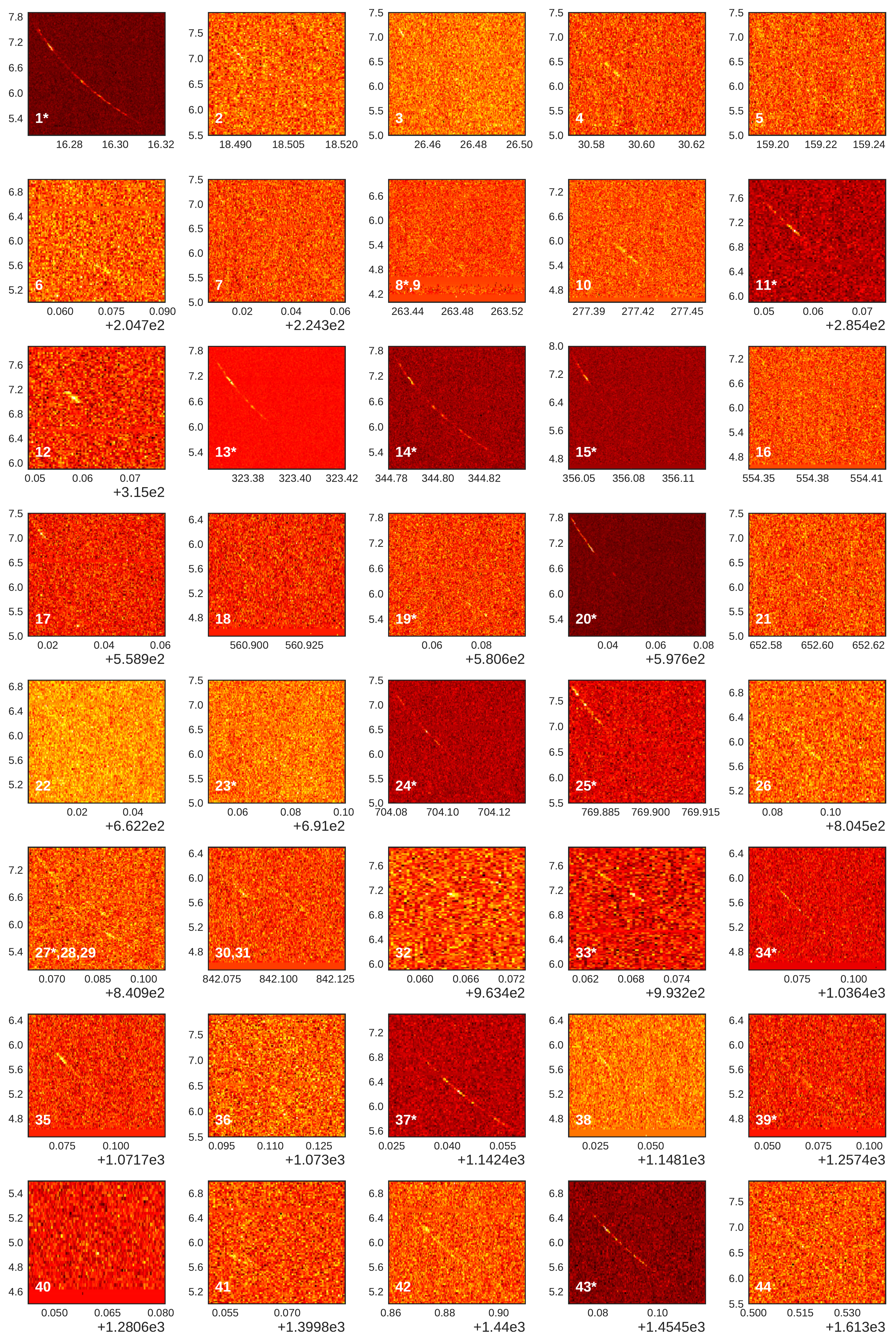}
\caption{Pulses detected from the first 30 minutes. The asterisks * indicate that the pulses have already been reported in \cite{vishal}. For all plots, time on the horizontal axes indicates seconds since the start of observation. Frequency on the vertical axes are in GHz. Numbering on the top-left corner of each panel corresponds to Table \ref{tab:param}. In cases when multiple pulses are shown in a panel, the numbering are in order of arrival time extrapolated to infinite frequency.  }
\label{fig:detections1}
\end{figure}

\begin{figure}
\includegraphics[width=\linewidth]{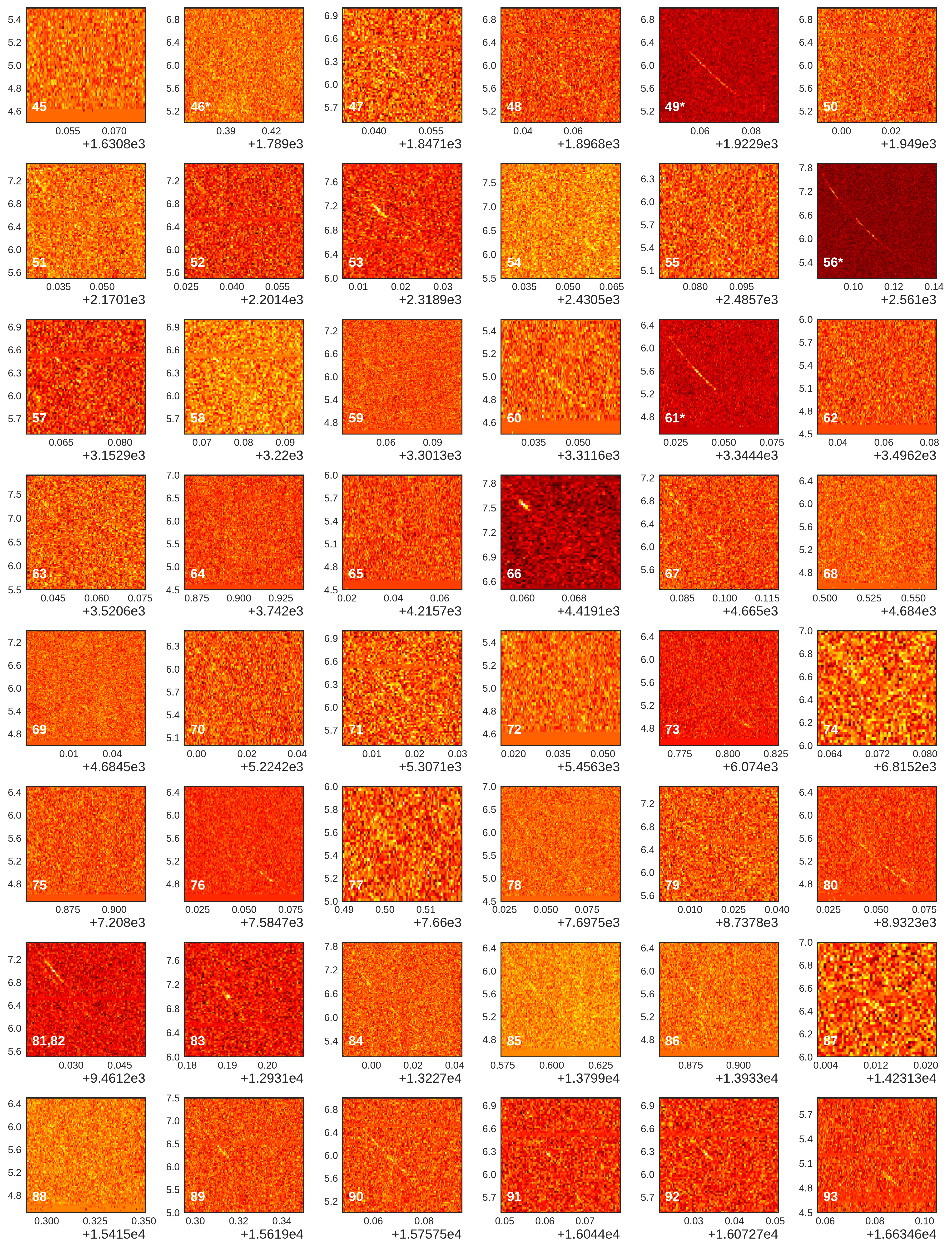}
\caption{Detected pulses (continued). Same as Fig. \ref{fig:detections1}. Note pulse morphologies vary and degree of visibility in the plot is subjective to frequency and time resolution shown.}
\label{fig:detections2}
\end{figure}

\begin{figure}
\includegraphics[width=1.1\linewidth]{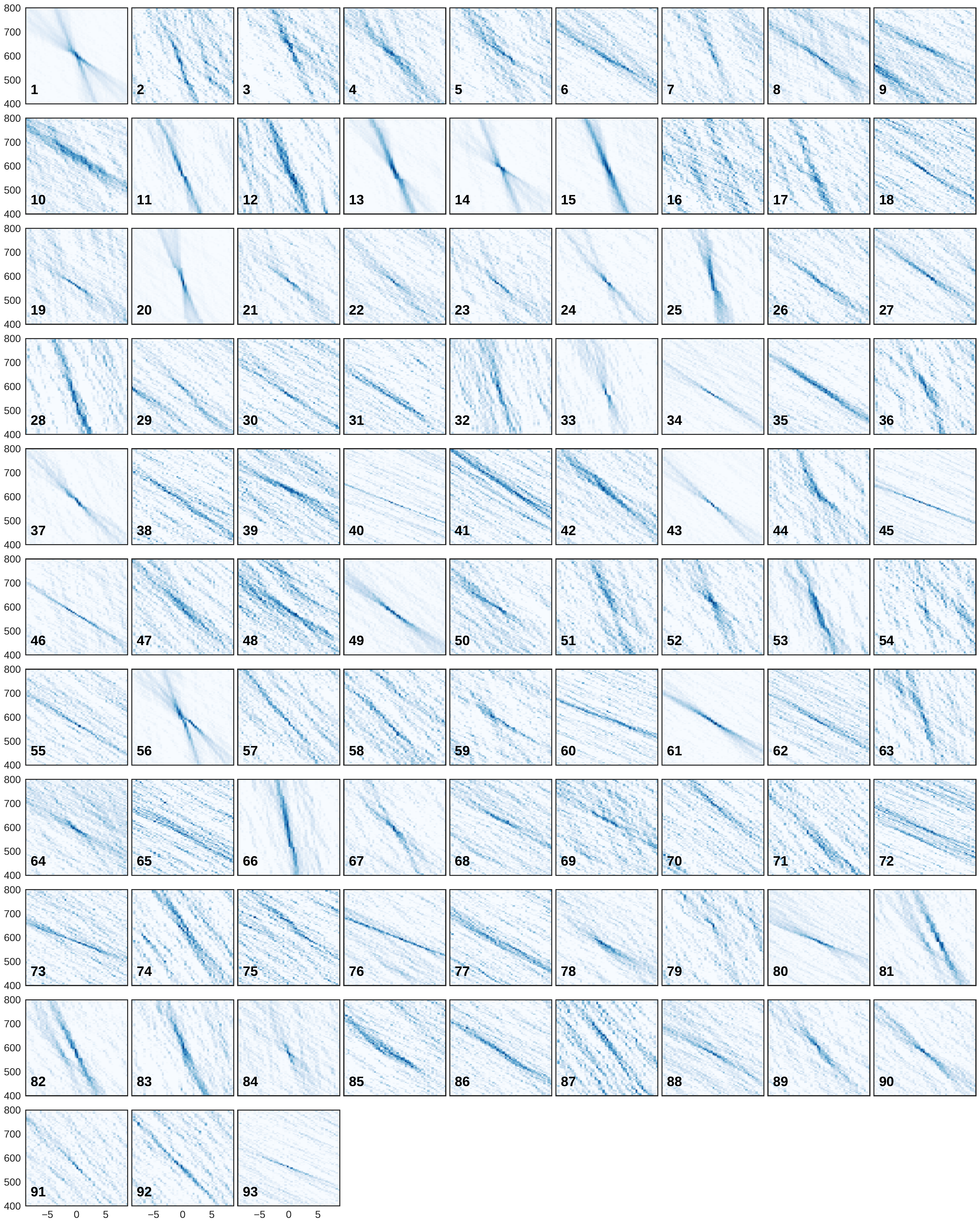}
\caption{Dedispersion verification plots of all pulses. All panels have the same scale. The horizontal axes are time of arrival extrapolated to the same frequency in milliseconds, and the vertical axes are DMs in the usual pc~cm$^{-3}$. The colors indicate flux density after incoherent dedispersion and the colors are scaled individually for each plot. A light color (high value) near the center of the panel indicate a close-to-expectation DM and TOA. }
\label{fig:butterfly}
\end{figure}

\begin{figure}
\includegraphics[width=1.1\linewidth]{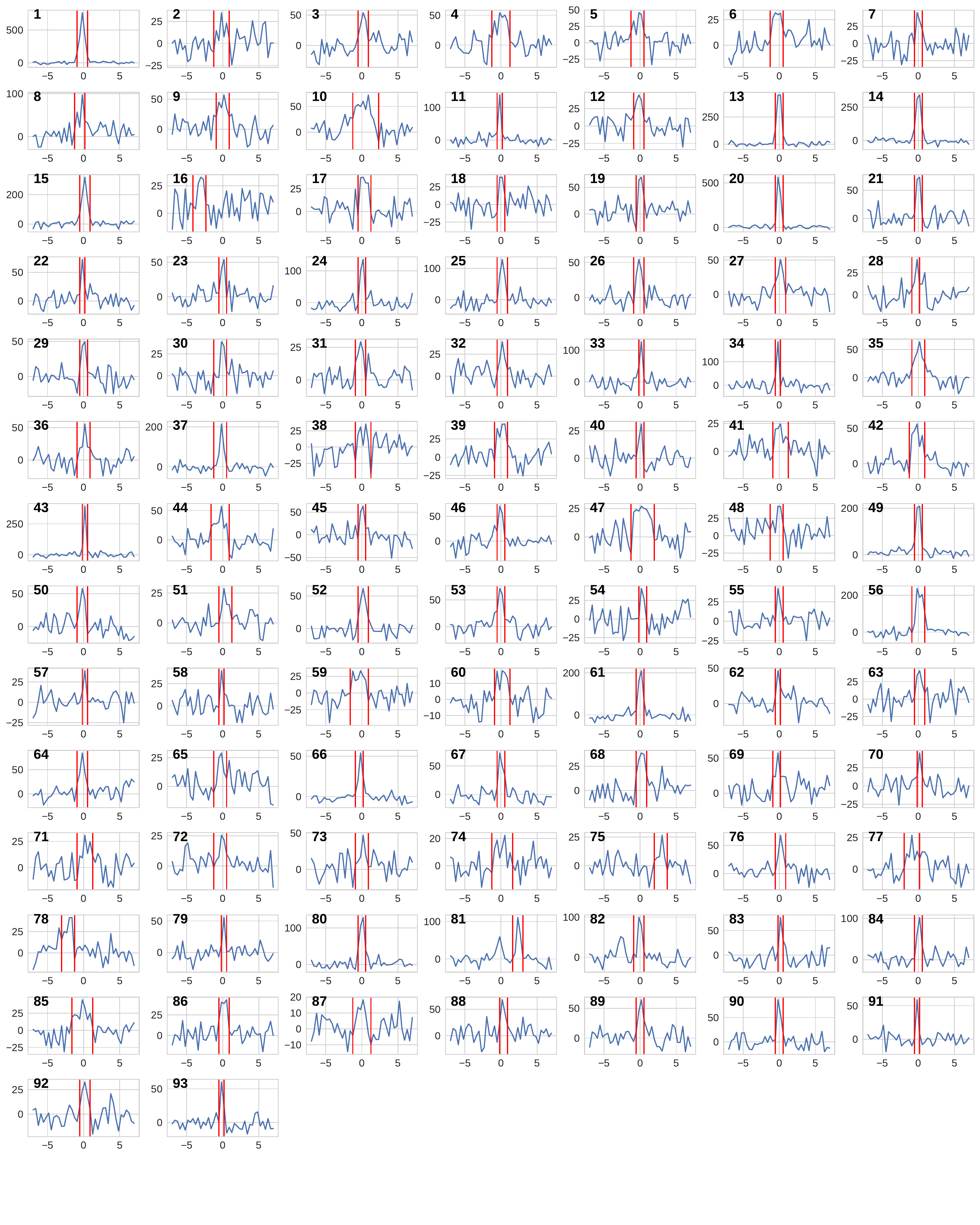}
\caption{Fluence determination for all pulses in fluence-maximizing sub-bands. Horizontal axes are time in milliseconds, while vertical axes are flux density in milli-Janskys. All pulses are dedispersed to individual S/N maximizing DM and in sub-bands indicated in Table \ref{tab:param}. Time resolutions are 0.35~ms. The time ranges used to compute the fluence are indicated with red vertical lines. }
\label{fig:timeseries}
\end{figure}

\newpage
\bibliographystyle{aasjournal}
\bibliography{references,CNN,blbibliography}

\end{document}